\documentclass[reprint,nofootinbib,preprintnumbers,showkeys,aps,physrev]{revtex4-2}
\usepackage{graphicx}
\usepackage{mathtools,amssymb} % mathtools loads amsmath
\usepackage{dcolumn} % Align table columns on the decimal point
\usepackage{bm} % bold math
\usepackage{xcolor}
\usepackage{dsfont} % double stroke for number sets
\usepackage{slashed} % Feynman/Dirac slash
\usepackage[hidelinks]{hyperref}

\newcommand \beq{\begin{eqnarray}}
\newcommand \eeq{\end{eqnarray}}

\newcommand{\R}{\mathds{R}}

\begin{document}
\unitlength=1mm
\allowdisplaybreaks

\title{Heavy-quark corner of the Columbia plot\texorpdfstring{\\}{} from the center-symmetric Curci-Ferrari model}

\author{V. Tomas \surname{Mari Surkau}}
\affiliation{Centre de Physique Th\'eorique, CNRS, Ecole polytechnique, IP Paris, F-91128 Palaiseau, France.}

\author{Urko Reinosa}
\affiliation{Centre de Physique Th\'eorique, CNRS, Ecole polytechnique, IP Paris, F-91128 Palaiseau, France.}

\date{\today}

\begin{abstract}
We investigate the heavy-quark corner of the Columbia plot using the gluon potential derived from the Curci-Ferrari extension of the Faddeev-Popov gauge fixing in the center-symmetric Landau gauge, as a proxy for the Polyakov loop potential. In line with the observation that Landau gauge couplings are not that large in the case of heavy-quark QCD, we consider a one-loop approximation and test its consistency by using various renormalization schemes while investigating the dependence of our results on the renormalization scale. We find that the main qualitative features of the phase structure in the heavy-quark regime are reproduced. Our results agree quantitatively with those of simulations to $10\%$ accuracy, which is the expected precision of the one-loop calculations in applications of the Curci-Ferrari model to Yang-Mills or heavy-quark QCD theories.
\end{abstract}

\maketitle

%%%%%
\section{Introduction}
Determining the phase structure of Quantum Chromodynamics (QCD) as a function of temperature and baryon chemical potential is the goal of many theoretical and experimental efforts in high-energy physics. It remains a key challenge in understanding the physics of the early universe and of extreme astrophysical objects \cite{Kolb2018TheUniverse, Weissenborn2011QUARKSTARS, Kurkela2014CONSTRAININGCHROMODYNAMICS, Borsanyi2016CalculationChromodynamics}. Accordingly, various heavy-ion collision experiments are devoted to resolving pressing questions about the QCD phase diagram \cite{Mohanty2011STARRHIC, ALICE2024TheQCD} at facilities such as the LHC or the RHIC.

On the theory side, it is pretty much accepted that asymptotic high-temperature and/or high-density regions are composed of deconfined matter, whose properties can be dealt with using methods based on perturbation theory \cite{Mogliacci:2013mca}. The study of the transition between hadronic matter and these deconfined phases, in contrast, requires the use of non-perturbative methods. In this respect, it is now well established by lattice simulations that, along the temperature axis, QCD presents a crossover transition \cite{Borsanyi2014FullFlavors, Bazavov2014EquationQCD} between two regimes characterized by different active degrees of freedom, hadrons on the one side, and quarks and gluons on the other side. Unfortunately, the extension of these simulations to the finite baryonic chemical potential case is seriously hindered by the ``sign problem'' that prevents the use of Monte-Carlo importance sampling \cite{Muroya2003LatticeReview, deForcrand2010SimulatingDensity}. Various workarounds, e.g., reweighting, Taylor expansion, and analytical continuation from imaginary chemical potentials, have been proposed \cite{Philipsen2012LatticePotential, Borsanyi2012QCD2, Nagata2022Finite-densityProblems}, but this prompts the exploration of complementary strategies. 

A very popular take is that of low-energy effective models that capture essential aspects of QCD, such as chiral symmetry breaking and the associated dynamical generation of mass \cite{Schaefer:1999em,Adhikari2018PionModel, Kovacs2016ExistenceModel}. More recently, these models have been extended by coupling them to the Polyakov loop potential that captures aspects related to the confinement/deconfinement transition \cite{Fukushima:2003fw,Ratti:2005jh,Roessner:2006xn}. These approaches are valuable in providing a qualitative picture of the QCD phase diagram. Yet, they usually involve many parameters, including the UV regulating function in cases where the model is not renormalizable, which certainly hinders their predictability \cite{Fujihara:2008ae}.

Another popular, and more first-principles, approach reformulates QCD as a tower of equations for its correlation functions. Different formulations exist, based, for instance, on Dyson-Schwinger equations \cite{Fischer2019QCDEquations} or the Functional Renormalization Group \cite{Fu2020QCDDensity}. As these methods rely on the calculation of primary correlation functions involving quarks and gluons, they require fixing the gauge and, very often, the Landau gauge is chosen for its special properties. In particular, it can be simulated on the lattice, allowing for a fruitful cross-fertilization between the methods. It is to be noted, however, that gauge fixing in the continuum is hindered by the Gribov copy problem \cite{Gribov1978QuantizationTheories, Singer1978SomeAmbiguity}. Thus, it is not fully established what starting gauge-fixed action should be used to write down the tower of equations for the correlation functions: the Faddeev-Popov action that results from ignoring the Gribov copy problem, or a yet-to-be-found extension of the Faddeev-Popov action that would take into account the Gribov copies. 

Another important aspect is that each possible tower of equations representing QCD contains infinitely many equations. Truncations are thus needed for any tractable prediction to be made. However, in a strongly coupled context such as QCD, it is not at all obvious how to find the appropriate truncations that keep the error under control. Enormous progress has been achieved recently in finding the appropriate truncations, not only based on feasibility but also, and more importantly, based on physical relevance \cite{Eichmann2016BaryonsStates, Huber2020NonperturbativeTheories}.

Over the past fifteen years, a third possible approach has been put forward based on an extension of the incomplete Faddeev-Popov action, as a way to take into account the Gribov copy problem phenomenologically. In a sense, this approach lies between the previous two, trying to reduce the number of phenomenological parameters and to improve error assessment. It departs from chiral effective models in that the matter sector is pure QCD, without phenomenological parameters, including the quark-gluon interaction. The only modeling occurs in the gauge sector, and more precisely in the extension of the Faddeev-Popov action, which now includes a gluon mass term. Interestingly, the corresponding Curci-Ferrari (CF) action \cite{Curci1976OnFields} is renormalizable. This implies that the approach contains one and only one phenomenological parameter, and becomes predictive once this parameter has been determined. The extension of the Faddeev-Popov action by a mass term is motivated by the decoupling behavior observed in Landau-gauge lattice simulations. In particular, the gluon propagator is found to saturate to a finite non-zero value in the infrared \cite{Bonnet2000InfraredLattice, Bonnet2001InfinitePropagator, Cucchieri2008ConstraintsTheories, Bogolubsky2009LatticeInfrared, Oliveira2012LatticeDependence, Maas2013GaugeTemperature, Silva2014GluonQCD}. As the lattice setup properly accounts for the Gribov copies, it is fair to consider that it can give some hints on what the Faddeev-Popov action could be missing.

The same simulations have revealed other interesting properties of quarks and gluons in the infrared that can help devise approximation schemes with good control over the error. In particular, although color fields are universally coupled at large momenta, it has been established that, for small Euclidean momenta, the interaction strength in the glue sector is up to four times smaller than the one in the matter sector \cite{Skullerud:2003qu}. Moreover, the interaction strength in the glue sector does not appear to be that large. This opens the possibility for a fascinating scenario in which QCD in the Landau gauge would be strongly coupled but with a weakly coupled glue
at its core that would help construct manageable and controlled approximations schemes.\footnote{It is to be stressed that these considerations come from Euclidean lattice simulations and, thus, apply in principle to applications of QCD in the Euclidean, of which the study of the QCD phase diagram is the main example.}

Interestingly enough, the CF model referred to above shares all the above-mentioned features with the lattice simulations and appears, thus, as an ideal playground to test the weakly coupled glue hypothesis. In particular, the CF model has been shown to reproduce the decoupling behavior of correlation functions obtained in Landau gauge simulations, \cite{Tissier2010InfraredTheory, Tissier2011InfraredCorrelators, Pelaez2013Three-pointTheory, Pelaez2014Two-pointGauge, Pelaez2015Quark-gluonModel, Reinosa2014Yang-MillsPerspective}. In the pure Yang-Mills case, or in the formal limit where all quarks are considered heavy, these results are obtained using simple one-loop perturbative calculations. The two-loop corrections are found to be tiny and to slightly improve the results, in agreement with the weakly coupled glue scenario.

Of course, perturbative methods cannot be used in the physical QCD case including light quarks. Yet, the weakly coupled glue scenario combined with the well-tested expansion in the inverse number of colors allows one to devise a controlled expansion scheme referred to as the rainbow-improved expansion scheme. The latter has been shown to reproduce the quark propagator quite accurately while capturing the spontaneous breaking of chiral symmetry \cite{Pelaez:2017bhh,Pelaez:2020ups}. Based on these premises, the approach has been applied to investigate QCD observables both in the vacuum and at finite temperature, see \cite{Pelaez2021AParameters} for a review. For instance, within the same controlled expansion, it was possible to evaluate the pion decay constant, with results in agreement with those of other approaches. At finite temperature, it was possible to investigate the phase structures of pure YM and heavy-QCD theories, with results in good agreement with those of lattice simulations. As for the physical QCD case, even though the approach is yet to be applied in its entirety in a finite-temperature setup, some preliminary results look quite promising regarding its capability to unveil some of the properties of QCD \cite{Maelger:2019cbk}.

It is to be noted that, in continuum calculations at finite temperature, it is necessary to generalize the Landau gauge fixing to the more general class of Landau-DeWitt or background Landau gauges \cite{DeWitt1967QuantumTheory}. The reason is that the center symmetry behind the confinement/deconfinement transition is not manifest within the Landau gauge fixing. The standard way to make this symmetry explicit within the class of background Landau gauges is to choose a self-consistent background, adjusted, at each temperature, to coincide with the gluon one-point function \cite{Polyakov1978ThermalLiberation, Braun2010QuarkConfinement, Reinosa2015DeconfinementTheory}. One benefit of this approach is that the self-consistent background is an order parameter for the confinement/deconfinement transition, and a proxy for the gauge-invariant but non-local order parameter, the Polyakov loop \cite{Polyakov1978ThermalLiberation}. This self-consistent (background) Landau gauge has also been implemented together with a CF mass term and has been found to capture the deconfinement transition in SU(2) and SU(3) YM theories as well as heavy-quark QCD through a background field potential whose minimum gives access to the self-consistent background \cite{Reinosa2015DeconfinementTheory, Reinosa2015PerturbativePotential, Reinosa2016Two-loopBeyond, Maelger2018PerturbativeCorrections}. It has been argued, however, that, because this potential does not correspond to the actual effective action, it can introduce unphysical biases in the presence of approximations. 

An alternative has been proposed that still leaves center symmetry manifest. It relies on the choice of a center-symmetric background \cite{vanEgmond2022ATransition,vanEgmond2024TheLattice}. The latter is not self-consistent and does not correspond to an order parameter. However, it can be shown that the one-point function in the presence of that background remains an order parameter.\footnote{In fact, it has been argued that, within this implementation, the gauge field correlators themselves become order parameters for the deconfinement transition \cite{vanEgmond2022ATransition}.} In addition, this order parameter results now from the minimization of a standard effective action, thus curing the limitation of the self-consistent backgrounds. The CF extension of this center-symmetric Landau gauge has been found to greatly improve the results obtained with the previous implementation of the Landau-DeWitt gauge in the case of pure YM theories \cite{vanEgmond2022ATransition,vanEgmond2023GaugeSymmetries, vanEgmond2024Center-symmetricConfinement}. In particular, the deconfinement temperature is obtained with great accuracy already at one-loop order and only shows a mild dependence on the chosen renormalization scheme or the renormalization scale, a signal of the good convergence properties of the approach \cite{MariSurkau2024DeconfinementDependences}. 

As a first step towards applications to QCD, this work extends the previous analyses to the heavy-quark QCD framework at finite temperature and density. The rich phase structure of this formal limit of QCD can be studied using simulations while avoiding the sign problem \cite{Fromm2012ThePotentials}. This then provides a valuable benchmark for continuum methods such as that based on the CF model. We show that all the known features of the top right corner of the Columbia plot \cite{Brown1990OnQuarks}, which represents the nature of the transition as a function of the quark masses, are correctly described, and perform a quantitative analysis of the results, assessing the dependence on both the renormalization scale and scheme, and comparing to simulations and background potential results.

The paper is organized as follows: Sec.~\ref{sec: framework} outlines the framework of the calculations, with a particular focus on the symmetries of the action and the Polyakov loop in heavy-quark QCD and their implications. In Sec.~\ref{sec: order params} we introduce the center-symmetric CF model and the one-loop effective potential for the gluon one-point function as a proxy for the Polyakov loop potential. Our results for vanishing, imaginary, and real chemical potentials are presented in  Sec.~\ref{sec: results}. Finally, we gather our conclusions in Sec.~\ref{sec: conclusions}, followed by some technical remarks about the symmetries of the potential in the appendices.

%%%%%
\section{Framework\label{sec: framework}}
In this work, we consider the Euclidean formulation of QCD as given by the action
\beq\label{eq: Eucl-QCD-action}
& & S[A,\psi_f,\bar\psi_f]\nonumber\\
& & \hspace{0.2cm}=\,\int_x\bigg\{\frac{1}{4}F_{\mu\nu}^aF_{\mu\nu}^a+\sum_{f=1}^{N_f}\bar{\psi}_f\left(\gamma_\mu{\cal D}_\mu+M_f-\mu\gamma_0\right)\psi_f\bigg\}.\nonumber\\
\eeq
The tensor $\smash{F_{\mu\nu}^a=\partial_\mu A_\nu^a-\partial_\nu A_\mu^a+gf^{abc}A_\mu^bA_\nu^c}$ is the non-Abelian field-strength tensor, with $g$ the coupling constant and $f^{abc}$ the structure constants of the SU(3) color group, while $\smash{{\cal D}_\mu=(\mathds{1}\partial_\mu-igA_\mu^at^a)}$ is the covariant derivative in the defining representation, with the associated generators $t^a$ taken such that $\smash{\operatorname{tr} t^at^b=\delta^{ab}/2}$. Summation over spinor and color indices is left implicit for the quark and antiquark fields, but we leave the summation over the various flavors explicit. We work with Euclidean Dirac matrices $\gamma_\mu$ which obey the anticommutation relations $\smash{\{\gamma_\mu,\gamma_\nu\}=2\delta_{\mu\nu}\mathds{1}}$ and which can be obtained from the usual Minkowskian ones as $\smash{\gamma_0=\gamma^0_M}$ and $\smash{\gamma_i=-i\gamma^i_M}$. 

From the Euclidean action \eqref{eq: Eucl-QCD-action}, one can evaluate the partition function
\beq
Z=\int{\cal D}[A\psi_f\bar\psi_f]\,e^{-S[A,\psi_f,\bar\psi_f]}\,,\label{eq: partition fct}
\eeq
and, more generally, thermal averages
\beq
\langle{\cal O}\rangle\equiv\frac{1}{Z}\int{\cal D}[A\psi_f\bar\psi_f]\,{\cal O}[A,\psi_f,\bar\psi_f]\,e^{-S[A,\psi_f,\bar\psi_f]}\,,\label{eq:O}
\eeq
which give access to the equilibrium properties of QCD. 

We recall that the temperature $T$ enters Eqs.~\eqref{eq: partition fct}-\eqref{eq:O} via the compactified temporal direction in Euclidean space. This means that the various fields are defined over the time interval $\smash{\tau\in [0,\beta]}$, with $\smash{\beta\equiv 1/T}$. The integration over Euclidean space corresponds to $\smash{\int_x\equiv\int_0^\beta d\tau\int d^{d-1}x}$. Moreover, the fields obey periodic or anti-periodic boundary conditions along the temporal direction:
\beq
A_\mu^a(\tau+\beta,\vec{x}) & = & A^a_\mu(\tau,\vec{x})\,,\label{eq:bc1}
\eeq
for the gluon field, and
\beq
\psi_f(\tau+\beta,\vec{x}) & = & -\psi_f(\tau,\vec{x})\,,\label{eq:bc2}\\
\bar\psi_f(\tau+\beta,\vec{x}) & = & -\bar\psi_f(\tau,\vec{x})\,,\label{eq:bc3}
\eeq
for the quark and antiquark fields. 

As for the chemical potential $\mu$, it multiplies the charge density\footnote{Note the different sign of $\mu$ as compared to the one used in Refs.~\cite{Reinosa2015PerturbativePotential, Maelger2018PerturbativeCorrections}. Our choice corresponds to $\smash{Z={\rm Tr}\,e^{-\beta(H-\mu Q)}}$, with $Q$ given in Eq.~\eqref{eq:charge}.}
\beq
n=\sum_{f=1}^{N_f}\bar\psi_f\gamma_0\psi_f\,,\label{eq:charge}
\eeq 
associated to a global U(1) symmetry of the QCD action and which characterizes the imbalance of quark and antiquark densities.\footnote{One could consider a separate chemical potential for each quark family, but for simplicity, we shall take them all equal.} The corresponding conserved charge $Q$ is actually related to the baryonic charge $\smash{Q_B=Q/3}$ whose chemical potential is $\smash{\mu_B=3\mu}$. We stress that a physical chemical potential is always real, but we shall also investigate the formal case of an imaginary chemical potential.

%%%
\subsection{Polyakov loop}
A particularly convenient thermal average that probes the thermodynamical properties of the system is the Polyakov loop\cite{Polyakov1978ThermalLiberation} defined as
\beq
\ell\equiv\langle\Phi\rangle\,,\label{eq:pl}
\eeq
with
\beq
\Phi[A]\equiv\frac{1}{3}{\rm tr}\,{\cal P}\,\exp\left\{ig\int_0^\beta \!\!d\tau A_0^a(\tau,\vec{x})t^a\right\},
\eeq
the normalized, time-ordered gluon field exponential. The relevance of $\ell$ lies in that it relates directly to the change $\Delta F_q$ in the system free energy upon inclusion of a static quark source:
\beq
\ell=e^{-\beta\Delta F_q}\,.\label{eq: l-DFq}
\eeq
A small Polyakov loop means that the energy cost for bringing such a static source into the system is very high, which one interprets as a confining phase. Instead, a Polyakov loop close to $1$ means that the energy cost is not that large, which one interprets as a non-confining phase. In the presence of a chemical potential that breaks the symmetry between quarks and antiquarks, one needs to distinguish between the Polyakov loop \eqref{eq:pl} and the anti-Polyakov loop defined as 
\beq
\bar\ell\equiv\langle\Phi^*\rangle\,,
\eeq 
which relates to the change $\Delta F_{\bar q}$ in the system free energy\footnote{At finite $\mu$, this is actually the Landau free energy or grand potential.} upon inclusion of a static antiquark source:
\beq
\bar\ell=e^{-\beta \Delta F_{\bar q}}\,.\label{eq: l-DFqb}
\eeq 
Of course, $\ell$ and $\bar\ell$ should coincide when $\smash{\mu=0}$.

Let us stress that, from the point of view of the Polyakov loop, a sharp distinction between a confining and a non-confining phase can be drawn only in the limit of infinitely heavy quarks. In this case, the theory benefits from a symmetry known as center symmetry, see below, which, when manifest, constrains the Polyakov loop to vanish and thus the free energy to diverge. In the presence of dynamical quarks, the physical interpretation of the Polyakov loop as relating to the free energy is still valid, but the center symmetry is explicitly broken from the start because the quark and antiquark boundary conditions \eqref{eq:bc2}-\eqref{eq:bc3} are not preserved by center transformations but turned instead into
\beq
\psi_f(\tau+\beta,\vec{x}) & = & -e^{-i2\pi k/3}\psi_f(\tau,\vec{x})\,,\label{eq:mbc}\\
\bar\psi_f(\tau+\beta,\vec{x}) & = & -e^{i2\pi k/3}\bar\psi_f(\tau,\vec{x})\,,\label{eq:mbcb}
\eeq
which differ from \eqref{eq:bc2}-\eqref{eq:bc3} when $k$ is not a multiple of $3$. For large enough quark masses, however, such as those that will be considered in this work, this explicit breaking has a small effect and the Polyakov loop still allows one to clearly distinguish between two phases, with either a small or a large free energy difference, and with a discontinuous transition between the two. As the quark masses are taken to lower values, this first-order transition turns into a second-order transition, defining a critical boundary in the space of quark masses, commonly referred to as the ``Columbia plot'' in the case of $2+1$ flavors. As explained in the Introduction, this paper aims to access this boundary line using alternative order parameters within the framework of the center-symmetric Landau gauge.

%%%
\subsection{Useful transformations}\label{sec:transfos}
The Euclidean QCD action benefits from several field transformations which, though not leaving the action invariant, connect it to the same action for different chemical potential values. Let us recall them here as they will come in handy later on.

A well-known example is that of charge conjugation, referred to in what follows as ${\cal C}$-transformation, that flips the sign of the chemical potential:
\beq
S_\mu[A^{\cal C},\psi_f^{\cal C},\bar\psi_f^{\cal C}]=S_{-\mu}[A,\psi_f,\bar\psi_f]\,.\label{eq:C}
\eeq
Another example is that of complex conjugation, referred to in what follows as ${\cal K}$-transformation, which complex conjugates both the chemical potential and the action:
\beq
S_\mu[A^{\cal K},\psi_f^{\cal K},\bar\psi_f^{\cal K}]=S_{\mu^*}[A,\psi_f,\bar\psi_f]^*\,.\label{eq:K}
\eeq
Yet another interesting example is that of particular abelian gauge transformations of the quark fields:
\beq
\psi_f^{{\cal A}_\alpha}(x)=e^{i\frac{\tau}{\beta}\alpha}\psi_f(x)\,, \quad \bar\psi_f^{{\cal A}_\alpha}(x)=e^{-i\frac{\tau}{\beta}\alpha}\bar\psi_f(x)\,,
\eeq
referred to in what follows as ${\cal A}_\alpha$-transformations, that shift the chemical potential by an arbitrary, purely imaginary amount:
\beq
S_\mu[A,\psi_f^{{\cal A}_\alpha},\bar\psi_f^{{\cal A}_\alpha}]=S_{\mu-i\alpha T}[A,\psi_f,\bar\psi_f]\,,\label{eq:A}
\eeq
with $\smash{\alpha\in\mathds{R}}$. As long as one considers only the transformation properties of the action, one can choose the parameter $\alpha$ to be an arbitrary real number. However, if one also wants to preserve the anti-periodic boundary conditions of the quark and antiquark fields, this parameter is constrained to be a multiple of $2\pi$. We shall refer to the corresponding transformations more simply as ${\cal A}$-transformations.

All the above transformations are not particular to QCD but are present in most theories involving charged fields in the presence of the corresponding chemical potentials. In the case of QCD at finite temperature, one can additionally consider center transformations. As we have mentioned above, these transformations leave the action invariant but modify the boundary conditions of the quark and antiquark fields from \eqref{eq:bc2}-\eqref{eq:bc3} to \eqref{eq:mbc}-\eqref{eq:mbcb}. Interestingly enough, though, one can restore the usual boundary conditions \eqref{eq:bc2}-\eqref{eq:bc3} by combining the center transformations with an ${\cal A}_{2\pi k/3}$-transformation. We refer to these combined transformations as ${\cal Z}$-transformations. They lead to the following transformation of the action
\beq
S_\mu[A^{\cal Z},\psi_f^{\cal Z},\bar\psi_f^{\cal Z}]=S_{\mu- i(2\pi k/3) T}[A,\psi_f,\bar\psi_f]\,.\label{eq:Z}
\eeq
We note that, by iterating these transformations, one can obtain \eqref{eq:A} with $\alpha$ a multiple of $2\pi$, which is then less fundamental than \eqref{eq:Z} in the case of QCD. We stress, however, that \eqref{eq:A} with $\alpha$ a multiple of $2\pi$ exists beyond the particular case of QCD, even when \eqref{eq:Z} does not apply. We shall build on this remark below when discussing the Roberge-Weiss transition.

The above transformations are relevant because they allow one to relate thermal averages at a given value of $\mu$ to similar thermal averages at another value of $\mu$. For the Polyakov loops in particular, the ${\cal C}$-transformation leads to
\beq
\ell(\mu)=\bar\ell(-\mu)\,,\label{eq:lC}
\eeq
while the ${\cal K}$-transformation leads to
\beq
\ell(\mu) & = & \ell^*(\mu^*)\,,\label{eq:lK1}\\
\bar\ell(\mu) & = & \bar\ell^*(\mu^*)\,.\label{eq:lK2}
\eeq
On the other hand, the ${\cal A}$-transformations lead to 
\beq
\ell(\mu) & = & \ell(\mu-i2\pi pT)\,,\label{eq:lA1}\\
\bar\ell(\mu) & = & \bar\ell(\mu-i2\pi pT)\,,\label{eq:lA2}
\eeq
while the ${\cal Z}$-transformations lead to
\beq
\ell(\mu) & = & e^{-ik2\pi/3}\ell(\mu-i(2\pi k/3)T)\,,\label{eq:lZ1}\\
\bar\ell(\mu) & = & e^{ik2\pi/3}\bar\ell(\mu-i(2\pi k/3)T)\,.\label{eq:lZ2}
\eeq

%%%%%
\subsection{Symmetry constraints}\label{sec:syms}
Of particular interest are those values of $\mu$ that are not changed by one of the above transformations. In this case, one obtains symmetry constraints on the Polyakov loops $\ell$ and $\bar\ell$. Before making these constraints more explicit, an important remark is in order.

We are here assuming that there is no spontaneously broken symmetry so that deep within each phase all thermal averages can be defined unambiguously, without the need to introduce an external source that is eventually sent to $0$. In this case, the symmetry constraints apply directly to the thermal averages. This remains true in the vicinity of a transition, particularly a first-order transition, because each spinodal branch is continuously connected to a stable branch deep within one phase. All the symmetry identities in this section are derived in this framework, as this situation certainly applies to any point in the interior of the Columbia plot. We shall later discuss how some of these identities need to be modified in the pure Yang-Mills case when the center symmetry is spontaneously broken.

After these words of caution, let us now analyze how the various symmetries constrain the Polyakov loops. For instance, a ${\cal C}$-transformation \eqref{eq:C} involves the change $\smash{\mu\to -\mu}$ which admits $\smash{\mu=0}$ as a fixed point. At this value of the chemical potential, from Eq.~\eqref{eq:lC}, we then have the constraint
\beq
\ell(\mu=0)=\bar\ell(\mu=0)\,.\label{eq:lC2}
\eeq
This is just the statement recalled above: in the absence of quark/antiquark asymmetry, it is impossible to distinguish them using an observable. 

We can apply a similar reasoning to the ${\cal K}$-transformation. The latter involves the change $\smash{\mu\to \mu^*}$ which admits any $\smash{\mu\in\mathds{R}}$ as a fixed point. Then, for any real chemical potential, from Eqs.~\eqref{eq:lK1}-\eqref{eq:lK2}, we find
\beq
\ell(\mu\in\mathds{R})\in\mathds{R} \quad {\rm and} \quad \bar\ell(\mu\in\mathds{R})\in\mathds{R}\,.
\eeq
These constraints are expected because they are necessary conditions for $\ell(\mu)$ and $\bar\ell(\mu)$ to write as \eqref{eq: l-DFq} and \eqref{eq: l-DFqb} with $\Delta F_q$ and $\Delta F_{\bar q}$ real.  

We can also combine ${\cal C}$ and ${\cal K}$. This leads to
\beq
\ell(\mu)=\bar\ell(-\mu^*)\,.\label{eq:lCK}
\eeq
This involves the change $\smash{\mu\to -\mu^*}$ which admits any $\smash{\mu\in i\mathds{R}}$ as a fixed point. Then, for any imaginary chemical potential,  this identity implies the constraint 
\beq
\bar\ell(\mu\in i\mathds{R})=\ell^*(\mu)\,,\label{eq:lim}
\eeq
which shows that, when switching from real to imaginary chemical potential, the number of real-valued, independent degrees of freedom is still two: from $\ell$ and $\bar\ell$ to $\smash{{\rm Re}\,\ell={\rm Re}\,\bar\ell}$ and $\smash{{\rm Im}\,\ell=-{\rm Im}\,\bar\ell}$.
 
As for the ${\cal Z}$-transformations, since they correspond to shifts of the chemical potential, a transformation that does not admit a fixed point, it seems that they do not lead to additional constraints at any given value of the chemical potential. However, we can again consider combinations with ${\cal C}$ or ${\cal K}$. Combining ${\cal Z}$ with ${\cal K}$ for instance leads to
\beq
\ell(\mu) & = & e^{-ik2\pi/3}\ell^*(\mu^*+i(2\pi k/3)T)\,,\label{eq:lZK1}\\
\bar\ell(\mu) & = & e^{ik2\pi/3}\bar\ell^*(\mu^*+i(2\pi k/3)T)\,.\label{eq:lZK2}
\eeq
This involves the change $\smash{\mu\to \mu^*+i(2\pi k/3)T}$ which admits $\smash{\mu=i(\pi k/3)T}$ as a fixed point. Then, for this value of the chemical potential, the above identities imply
\beq
\ell(i(\pi k/3)T) & = & e^{-i2\pi k/3}\ell^*(i(\pi k/3)T)\,,\label{eq:fp1}\\
\bar\ell(i(\pi k/3)T) & = & e^{i2\pi k/3}\bar\ell^*(i(\pi k/3)T)\,,\label{eq:fp2}
\eeq
which fix the phase of $\ell(i(\pi k/3)T)$ to $-\pi k/3$  modulo $\pi$ and also, in agreement with Eq.~\eqref{eq:lim}, the phase of $\bar\ell(i(\pi k/3)T)$ to $\pi k/3$  modulo $\pi$. 

Various remarks are in order at this point. First, the identities \eqref{eq:fp1} and \eqref{eq:fp2} correspond to symmetry constraints in the sense that they derive from a symmetry of the action. That this is so is not completely obvious because $S_{i(\pi/3)T}[A,\psi_f,\bar\psi_f]$ is not left invariant by the transformation ${\cal KZ}$, but, rather, changed into its complex conjugate $S_{i(\pi/3)T}[A,\psi_f,\bar\psi_f]^*$. To see that ${\cal KZ}$ is indeed a symmetry for $\smash{\mu=i(\pi k/3)T}$, it is convenient to express the Polyakov loop as a functional integral over the gluon field only, upon exact integration of the quark and antiquark fields. One finds
\beq
\ell=\frac{1}{Z}\int{\cal D}A\,\Delta[A;\mu]\,\Phi[A]\,e^{-S_{YM}[A]}\,,
\eeq
where $\Delta[A;\mu]$ is the well known fermionic determinant. The latter obeys the property
\beq
\Delta_\mu[A]=(\Delta_{-\mu^*}[A])^*
\eeq
which makes it real for any imaginary chemical potential. One can even show that the fermionic determinant is positive in this case. Using these observations and coming back to the transformation ${\cal KZ}$, we find that $S_\mu[A]\equiv S_{\rm YM}[A]+\ln\Delta_\mu[A]$ is invariant under ${\cal KZ}$ when $\smash{\mu=i(\pi k/3)T}$.

The second remark is that even though the identities \eqref{eq:fp1} and \eqref{eq:fp2} correspond to symmetry constraints, they have a very different status depending on the parity of $k$. In particular, as we now discuss, whether or not these identities can be broken spontaneously, depends on the parity of $k$.  

Let us first consider the case of an even $k$. Because ${\cal Z}$ is never a symmetry of the action in the presence of quarks, one can restrict to the case $\smash{k=0}$. Indeed, whatever happens there, that is whether or not the corresponding constraint is fulfilled, will be mapped to the other even values of $k$ upon the use of Eqs.~\eqref{eq:lZ1}-\eqref{eq:lZ2}. But now, the symmetry constraint at $\smash{k=0}$ is $\smash{\ell(0)=\ell^*(0)}$, that is the statement that $\ell(0)$ should be real. Because this is a necessary condition for $\ell(0)$ to be interpreted as a free energy, we expect the associated symmetry ${\cal K}$ to never break spontaneously. Upon using Eqs.~\eqref{eq:lZ1}-\eqref{eq:lZ2}, this also implies that the phase of $\ell(\pm i(2\pi/3)T)$ should always be $\mp 2\pi/3$ modulo $\pi$.

Let us now consider the odd $k$ case. As before, one can restrict to one particular value of $k$ and deduce the other odd values upon application of Eqs.~\eqref{eq:lZ1}-\eqref{eq:lZ2}. One could think of choosing $\smash{k=1}$ but a more convenient choice is $\smash{k=3}$ and thus $\smash{\mu=i\pi T}$. The reason is that, in this case, the transformation ${\cal KZ}$ is nothing but ${\cal KA}$. One could first think that this symmetry can never break spontaneously because, as we have just argued, ${\cal K}$ cannot break spontaneously, and neither can ${\cal A}$ as it corresponds to a gauge transformation. In fact, the argumentation would be correct if ${\cal K}$ and ${\cal A}$ were symmetries of the action for $\smash{\mu=i\pi T}$. In this case, the spontaneous breaking of ${\cal KA}$ would imply that of ${\cal K}$ or ${\cal A}$ and thus a contradiction. However, for $\smash{\mu=i\pi T}$, neither ${\cal K}$ nor ${\cal A}$ are symmetries of the action. In this case, ${\cal KA}$ could break spontaneously without this leading to any contradiction because ${\cal K}$ and ${\cal A}$ are already explicitly broken. The spontaneous breaking of ${\cal KA}$ does occur above some temperature and is known as the Roberge-Weiss transition. It is characterized by the fact that, unlike $\ell(0)$, which is always real,  $\ell(i\pi T)$ is not necessarily real as the symmetry ${\cal KA}$ would imply. Upon using Eqs.~\eqref{eq:lZ1}-\eqref{eq:lZ2}, this also implies that the phase of $\ell(\pm i(\pi/3)T)$ is not necessarily $\mp\pi/3$ modulo $\pi$ as the symmetry ${\cal KZ}$ would imply.

%%%
\subsection{Additional remarks}\label{sec:rmqs}
The Roberge-Weiss transition is generally associated with center symmetry. What the above discussion for odd $k$ shows, however, is that it is actually rooted in the combination ${\cal K A}$ which has a priori nothing to do with center symmetry. The role of center symmetry is to map what happens at $\smash{\mu=i\pi T}$ to $\smash{\mu=\pm i(\pi/3)T}$. So, whether the symmetry constraints are fulfilled or not at $\smash{\mu=\pm i(\pi/3)T}$ has to do with the explicit realization or spontaneous breaking of ${\cal KA}$. This opens the possibility for Roberge-Weiss-type transitions to exist in theories that do not have center symmetry but obey the more general ${\cal KA}$ symmetry.

The discussion for even $k$ is also illuminating as it is closely related to the spontaneous breaking of center symmetry in pure Yang-Mills theories. In such theories, center symmetry is an actual symmetry of the action that can break spontaneously. As we have recalled, the broken phase is characterized by a non-vanishing Polyakov loop $\ell$, interpreted in terms of a finite free energy $\Delta F_q$ from Eq.~\eqref{eq: l-DFq}. There seems to be a problem, however, because, as is usually the case when a symmetry is spontaneously broken, there is not a unique possible value of the order parameter, but various ones, related by the symmetry transformations, here multiplications by $e^{i2\pi/3}$ or $e^{-i2\pi/3}$. This means that there are at least two broken realizations of the Polyakov loop which are not real, and, as such, cannot be interpreted in terms of a free energy. Two questions emerge then: is one of the broken realizations of the Polyakov loop compatible with the free energy interpretation? And what is the interpretation of the other two broken realizations?

To answer these questions, one should recall that the proper description of spontaneous symmetry breaking requires introducing a small perturbation that breaks the symmetry explicitly, and then, studying the fate of the symmetry as this perturbation is made smaller and smaller. The spontaneously broken symmetry corresponds to the case where the breaking persists, even after the perturbation has been removed. Since the Polyakov loops are gauge invariant, this symmetry-breaking perturbation should also be gauge-invariant. In fact, we already know how to introduce such a perturbation: we just need to include quarks with a finite mass $\smash{M_f=M}$. 

This represents an explicit breaking of center symmetry because the quark and antiquark anti-periodic boundary conditions \eqref{eq:bc2}-\eqref{eq:bc3} are modified to \eqref{eq:mbc}-\eqref{eq:mbcb} under center transformations. Because of that, the symmetry does not constrain the value of the Polyakov anymore, but rather connects the values of the Polyakov loop over different types of boundary conditions for the quarks:
\beq
\langle\Phi[A]\rangle=e^{-i2\pi/3}\langle\Phi[A]\rangle_{e^{-i2\pi/3}}\,,
\eeq
where $\langle\dots\rangle_{e^{\pm i2\pi/3}}$ refers to the average over quark and antiquark fields obeying the modified boundary conditions. In principle, one would expect that, as one tries to remove the quarks by making them heavier and heavier ($M\to \infty$), the above averages become insensitive to the boundary conditions of the quarks. In this case, the above relation turns into a constraint
\beq
\langle\Phi[A]\rangle=e^{-i2\pi/3}\langle\Phi[A]\rangle\,,
\eeq
that fixes the Polyakov loop to be $0$. However, suppose now that the averages keep a memory of the quark boundary conditions despite the limit $M\to\infty$. In that case, there is no constraint on the Polyakov loop, and the above equation relates the Polyakov loop in different broken symmetry sectors. Since the modified boundary conditions can be re-interpreted in terms of imaginary chemical potentials, see above, this establishes the connection between the spontaneous breaking of center symmetry in the YM case and the discussion for even values of $k$ given above.
  
As a final remark, let us return to the real-valuedness of $\ell(0)$ and $\ell(i\pi T)$. As we have already explained, $\ell(0)$ should always be real, while $\ell(i\pi T)$ is real as long as the Roberge-Weiss symmetry is not broken. For $\smash{\mu=0}$, $\ell(0)$ better be positive for the free energy interpretation to hold. This is what is usually seen in the simulations, and our results to be presented below will also yield a positive $\ell(0)$. However, proving this statement from the definition of $\ell$ as a thermal average is more difficult than it first meets the eye. One could imagine invoking the fact that the fermionic determinant is positive in this case, but the latter is also positive for $\smash{\mu=i\pi T}$ while $\ell(i\pi T)$ will be found to be negative below.

%%%%%
\section{Alternative order parameters\label{sec: order params}}
So far, we have reviewed mostly known material in terms of the Polyakov loops $\ell$ and $\bar\ell$. These order parameters are non-local constructs, however, which makes their direct evaluation in the continuum hard.\footnote{The situation is far more favorable on the lattice where the Polyakov loops appear as a product of gauge links along the temporal direction.} It is then worth asking whether one can construct alternative order parameters for the confinement/deconfinement transition from simpler quantities such as, for instance, gauge correlators.

The gauge correlators are gauge-dependent quantities, however, and it is then not obvious how they could be used to probe a physical phase transition. This question was recently addressed in Refs.~\cite{vanEgmond2023GaugeSymmetries, vanEgmond2024Center-symmetricConfinement} where it was shown that, in some appropriately chosen gauges, the gauge-field correlators become order parameters for center symmetry and, thus, probes for the confinement/deconfinement transition. The gauge dependence of these objects just means that the very same correlators computed in other gauges have no reason to be order parameters. They only play this role in those specific gauges, dubbed as center-symmetric.

%%%
\subsection{The center-symmetric Landau gauge}
One particular example of a center-symmetric gauge is the center-symmetric Landau gauge, a particular realization of the family of background Landau gauges
\beq\label{eq: gauge-fix-condition}
    0=\bar{D}_\mu^{ab}(A_\mu^b-\bar{A}_\mu^b)\,,
\eeq
where $\bar A_\mu^a$ is a background gauge-field configuration specifying the gauge and $\smash{\bar D_\mu^{ac}\equiv \partial_\mu\delta^{ac}+gf^{abc}\bar A_\mu^b}$ is the adjoint covariant derivative for that background. It is shown in Refs.~\cite{vanEgmond2022ATransition, vanEgmond2024Center-symmetricConfinement}, that the gluon one-point function
\beq
\langle A_\mu^a\rangle\equiv\frac{\int{\cal D}_{\rm gf}[A\psi_f\bar\psi_f]\,A_\mu^a\,e^{-S[A,\psi_f,\bar\psi_f]}}{\int{\cal D}_{\rm gf}[A\psi_f\bar\psi_f]\,e^{-S[A,\psi_f,\bar\psi_f]}}\label{eq:one_point}
\eeq 
becomes an order parameter for center symmetry, and thus a proxy for the Polyakov loop, when the background is chosen in some specific configurations, dubbed as center symmetric. The notation ${\cal D}_{\rm gf}[A\psi_f\bar\psi_f]$ used in Eq.~\eqref{eq:one_point} represents the gauge-fixed measure in that specific gauge.

For simplicity, and without loss of generality, the background field can be chosen in the subspace of configurations that explicitly satisfy the symmetries of the system at finite temperature, which for a medium in thermal equilibrium means homogeneity and isotropy. In particular, one restricts to temporal and homogeneous backgrounds. The latter can, in addition, always be color-rotated into the Cartan subalgebra,
\begin{equation}\label{eq: bkgrd field def}
    \bar{A}^a_\mu(x)=\frac{T}{g}\delta_{\mu0}\delta^{aj}\bar r^j\,,
\end{equation}
where the label $j$ can take values along the two Abelian directions of the SU(3) algebra, $\smash{j=3}$ or $8$, corresponding to the two diagonal Gell-Mann matrices $\lambda^3$, $\lambda^8$. 

The real numbers $\bar{r}^j$ are the components of a constant vector $\smash{\bar{r}=(\bar r^3,\bar r^8)\in\R^2}$ with mass dimension $0$.  The center-invariant backgrounds correspond to particular instances of this vector, denoted $\bar r_c$. They are found using the notion of Weyl chambers. These are (equilateral) triangular regions, see Fig.~\ref{fig: mu_i min trajectories} below, paving the plane of the vector $\bar r$, or the plane of the vector $r$ corresponding to the gluon one-point function, which we will introduce in Sec.~\ref{subsec: gluon avg potential}. They are connected by reflections about their edges, which represent gauge transformations. Center transformations, on the other hand, appear as translations of the Weyl chambers along their edges, but, thanks to the gauge transformations, these can be equivalently recast into rotations of the Weyl chambers by an angle $\pm 2\pi/3$ about their centers of mass(centroids). This means that the centroid of each Weyl chamber is a center-symmetric background. 
One example that we shall consider in this work is $\smash{\bar r_c=(4\pi/3,0)}$. Any other choice would also allow one to use the one-point function \eqref{eq:one_point} as an order parameter for center symmetry, leading to identical physical results. We point to Ref.~\cite{vanEgmond2024Center-symmetricConfinement} for further details.

%%%
\subsection{The center-symmetric Curci-Ferrari model}
The textbook procedure to take into account the gauge-fixing condition \eqref{eq: gauge-fix-condition} is to use 
\beq
{\cal D}_{\rm gf}[A\psi_f\bar\psi_f]={\cal D}[A\psi_f\bar\psi_f]\int {\cal D}[c,\bar c,h]\,e^{-\delta S_{\rm FP}[A,c,\bar c,h]}\,,\label{eq:gf}\nonumber\\
\eeq
as the gauge-fixing measure in Eq.~\eqref{eq:one_point}, with
\begin{equation}\label{eq: fixing-term action}
\delta S_{\rm FP}=\int_x\Big\{\big(\bar D_\mu\bar c\big)^a \big(D_\mu c\big)^a+ih^a\big(\bar D_\mu (A_\mu-\bar A_\mu) \big)^a\Big\}\,,
\end{equation}
the Faddeev-Popov (FP) gauge-fixing terms, where $c^a$, $\bar c^a$ and $h^a$ denote the standard ghost, antighost and Nakanishi-Lautrup fields. When the background is taken center-symmetric, the gauge-fixed Yang-Mills action $S_{YM}+\delta S_{\rm FP}$ becomes center-invariant and allows one to derive constraints on the one-point function \eqref{eq:one_point} as well as on higher correlators \cite{vanEgmond2024Center-symmetricConfinement}, turning them into potential order parameters for center symmetry.

It should be stressed, however, that the FP procedure leading to \eqref{eq:gf}-\eqref{eq: fixing-term action} suffers from a serious loophole as it assumes the absence of Gribov copies in the family of background Landau gauges, an assumption known to be incorrect. At high energies, the Gribov copies are not expected to play a major role, and thus the FP procedure is believed to be a sensible approach. But this is not necessarily so at low energies, in particular regarding the low-temperature phase of the system. In this case, the Faddeev-Popov action likely needs to be modified to take the Gribov copies into account properly.

To date, there is no known way to implement this program consistently in the continuum. Very promising approaches exist on the market (such as the Gribov-Zwanziger framework or its various refinements \cite{Zwanziger1989LocalHorizon, Dudal2008RefinementResults}), but they only partially deal with the problem. A more phenomenologically inspired alternative consists of modeling and constraining the terms that could be missing in Eq.~\eqref{eq:gf}.

In this respect, one possible source of constraints comes from the lattice implementation of the background Landau gauges.\footnote{These gauges can be formulated as an extremization problem, which, if restricted to minimization or maximization, can be implemented numerically. This is the standard way the Landau gauge is simulated on the lattice. For a lattice implementation of the center-symmetric Landau gauge considered here, see Ref.~\cite{vanEgmond2024TheLattice}.} The lattice can choose one copy per gauge orbit, so, even though there remains an ambiguity in the sense that there are as many possible ways of fixing the gauge as there are ways to choose the copies, each choice is a rigorous one, unlike what happens in the Faddeev-Popov implementation. In turn, the lattice is an important source of information concerning the looked-after extension of the Faddeev-Popov action.

In the case of the Landau gauge for instance, corresponding to the choice $\smash{\bar A=0}$, the lattice simulations have shown that, while the ghost propagator remains essentially unchanged in the infrared compared to its tree-level counterpart, the gluon propagator experiences a drastic change from a tree-level massless propagator to an infrared screened one. This breaking of scale invariance, known as decoupling, is believed to be connected to the dynamical generation of mass in non-gauge-fixed YM theories whose manifestation within the Landau gauge could involve different mechanisms, such as the Schwinger mechanism, see e.g. \cite{Fischer2019QCDEquations} for a review. Let us stress, however, that these mechanisms are usually formulated within the Faddeev-Popov approach, which, as we have just explained, needs to be extended to cope with the Gribov copy problem. 

In fact, the very choice of copies along each orbit is also a source of (explicit) scale invariance breaking.\footnote{Indeed, the Landau gauge condition is invariant under scale transformations $A_\mu(x)\to \lambda A_\mu(x/\lambda)$ and this implies that the formal ensemble of all gauge configurations (including all copies) obeying $\partial_\mu A_\mu^a=0$ is invariant under scale transformations. However, the very operation of selecting one copy per gauge orbit, to create the ensemble with which the functional integral is evaluated eventually, breaks this scale invariance.} Although not as fundamental as the physical breaking, as it would not show up in observables, this second source of breaking can affect the precise details of the correlation functions in the infrared. Stated differently, the very operation of selecting copies contains hidden gauge-fixing parameters that affect the correlation functions \cite{Maas2008MoreTheory, Maas2011OnOrbit, Maas2011PropertiesOrbits}. Knowing these hidden parameters exactly for a given choice of copies is a tremendously hard task. 

The phenomenological approach referred to above builds on these observations and proposes an extension of the FP action relying on the Curci-Ferrari model, which consists in the addition of a gluon mass term $m^2(A_\mu^a)^2/2$ rooted in the observed decoupling behavior. One very welcome feature of the model is that it is renormalizable. This means that, despite the presence of one more parameter that needs to be fixed in addition to the gauge coupling $g$ (thus further emphasizing its phenomenological nature), the renormalization of the loop corrections does not require introducing even further parameters. In other words, once the mass parameter is fixed, the model becomes predictive. In a certain sense, this extra parameter can be understood as the product $\smash{m=\xi m_{\rm phys}}$ of the physical mass scale $m_{\rm phys}$ times a dimensionless parameter $\xi$ that takes effectively into account the hidden gauge-fixing parameters referred to above. Of course, the physical mass cannot be predicted within the CF model, so the fitting of $m$ should rather be understood as a fitting of $\xi$ as a way to approximate as well as possible the hidden details of the gauge-fixing procedure while incorporating phenomenologically the dynamically generated mass.

Beyond these formal considerations, the CF model has shown a surprising ability to capture many infrared properties of Landau gauge QCD and YM theories, see Ref.~\cite{Pelaez2021AParameters} for a thorough review. Moreover, in the pure YM case (and also in the heavy-quark regime), these properties can be captured from a simple perturbative approach, in line with the fact that the lattice simulations lead in this case to a perturbative expansion parameter $\lambda=g^2N_c/16\pi^2<1$.

As for the center-symmetric Landau gauge, it has only recently been implemented on the lattice, see Ref.~\cite{vanEgmond2024TheLattice}. In a work in preparation, it will be shown that the lattice gluon propagator in this gauge is also screened. Therefore, it makes sense to model the extension beyond the FP terms by a mass term. To make sure that the extended action is center-symmetric when the background is taken center-symmetric, we choose the mass term of the form
\begin{equation}\label{eq: CF action}
 \delta S_{CF}=\int_x\frac{1}{2}m^2\big(A_\mu^a-\bar A_\mu^a\big)^2\,.
\end{equation}
Interestingly, because the center-symmetric background vanishes at low temperature, due to the explicit factor of $T$ in Eq.~\eqref{eq: bkgrd field def}, by choosing a zero-temperature renormalization scheme, we can use for $m$ the same parameter as the one in the CF extension of the Landau gauge FP action, which is fixed by fitting the Landau gauge lattice propagators. 

Below, to study the stability of our results, we shall consider two popular schemes within the CF framework corresponding to two possible definitions of the renormalized mass at zero temperature: the vanishing momentum (VM) scheme and the infrared-safe (IRS) scheme. Of course, the physical results should not depend on the chosen scheme and we shall use this as a test of the quality of our approach/approximations.

%%%
\subsection{The gluon average potential\label{subsec: gluon avg potential}}
Our goal is then to study the gluon one-point function \eqref{eq:one_point} in the center-symmetric Landau gauge \eqref{eq: gauge-fix-condition} with a center-symmetric background \eqref{eq: bkgrd field def}, phenomenologically completed in the infrared using the CF model \eqref{eq: CF action}, and to use it as a probe of the QCD phase structure in the regime of heavy quarks. 

It can be shown that, for the here considered background, and more generally for backgrounds of the form \eqref{eq: bkgrd field def}, the one-point function takes the form
\begin{equation}\label{eq: form of gauge expval}
    \langle A^a_\mu(x)\rangle=\frac{T}{g}\delta_{\mu0}\delta^{aj}r^j\,,
\end{equation}
with $r^j$ depending on the temperature and the chemical potential. This means that, to evaluate the one-point function, we need only to consider the effective potential for $\smash{r=(r^3,r^8)}$ whose extremization will give the actual values of $r^3$ and $r^8$ at each temperature and chemical potential. Moreover, when the background is chosen center-symmetric, $\smash{\bar r=\bar r_c}$, then the one-point function and thus $r$ becomes an order parameter for center symmetry in the sense that it should equal $\bar r_c$ as long as the symmetry is manifest, and any deviation from $\bar r_c$ will signal the breaking of the symmetry.

At one-loop order, the potential for any choice of background $\bar r$ and any value of $r$ appears as the sum of a glue contribution and a matter contribution:
\beq
V_{\bar r}(r)=V^{\rm glue}_{\bar r}(r)+V^{\rm matter}(r)\,.
\eeq
\vglue2mm
In Ref.~\cite{MariSurkau2024DeconfinementDependences}, the glue contribution was computed to be\hspace*{-.5cm} 
\begin{widetext}
    \beq
        V^{\rm glue}_{\bar r}(r) 
        & = & \frac{3T}{2\pi^2}\sum_\kappa\int_0^\infty \!\!dq\,q^2{\rm Re}\,\ln\Big[1-e^{\varepsilon_q/T-i\kappa\cdot\bar{r}}\Big]-\frac{T}{2\pi^2}\sum_\kappa\int_0^\infty \!\! dq\,q^2{\rm Re}\,\ln\Big[1-e^{q-i\kappa\cdot\bar{r}}\Big]\nonumber\\
        & + & \frac{T}{2\pi^2}\sum_\kappa (\kappa\cdot\Delta r) \int_0^\infty \!\!\!dq\,q^2\,{\rm Im}\,\Big[3n_{\varepsilon_q-i\kappa\cdot\bar{r}T}-n_{q-i\kappa\cdot\bar{r}T}\Big]+\frac{T^4}{6\pi}\sum_\kappa(\kappa\cdot\Delta r)^3\left(\frac{\kappa\cdot\bar r}{\pi}-1+\frac{\kappa\cdot\Delta r}{4\pi}\right)\nonumber\\
        & + & \frac{T^2}{2}(\Delta r)^2\left[\frac{Z_a Z_{m^2}}{g^2}+\frac{3N_c}{64\pi^2}\left(\frac{1}{\epsilon}+\ln\frac{\bar\Lambda^2}{m^2}+\frac{5}{6}\right)\right]m^2 \nonumber\\
        & + & \frac{T^2}{4\pi^2}\sum_\kappa(\kappa\cdot\Delta r)^2\int_0^\infty dq\,q^2\,{\rm Re}\,\left[\left(3\frac{m^2}{q^2}+6+\frac{q^2}{m^2}\right)\frac{n_{\varepsilon_q-i\kappa\cdot\bar{r}T}}{\varepsilon_q}-\left(\frac{1}{2}+\frac{q^2}{m^2}\right)\frac{n_{q-i\kappa\cdot\bar{r}T}}{q}\right]\nonumber\\
        & + & \frac{T}{\pi}\sum_\kappa\sum_{q\in\mathds{Z}}\bigg[\frac{1}{12}\Big( 2\,{\rm Re}\,X_0^3 + 3\bar X^3 - \bar X_0^3 - 2X^3 - X_+^3 - X_-^3 \Big)-\,\frac{T}{4}(\kappa\cdot\Delta r)\,\bar\omega_q^\kappa\big( \bar X_0-3\bar X \big)\nonumber\\
        & & \hspace{1.0cm}   +\,\frac{T^2}{4}(\kappa\cdot\Delta r)^2\bar X+\frac{T^3}{6}(\kappa\cdot\Delta r)^3{\rm sgn}(\bar\omega^\kappa_q) +\,\frac{T^2}{16}(\kappa\cdot\Delta r)^2(\bar\omega_q^\kappa)^2\left(\frac{6}{\bar X}-\frac{1}{\bar X_0}-\frac{2}{\bar X_0 + \bar X}\right)\bigg]\,,\label{eq: Vglu}
    \eeq
\end{widetext}
where we have introduced the short-hand notations $\smash{\Delta r\equiv r-\bar r}$ and $\smash{n_\varepsilon=1/(e^{\beta\varepsilon}-1)}$ with $\smash{\varepsilon_q=\sqrt{q^2+m^2}}$, and the summation symbol represents a sum over the adjoint weights, which play the role of adjoint color indices in an appropriate Cartan-Weyl basis. The last two lines of Eq.~\eqref{eq: Vglu} are convergent Matsubara sums that have to be performed numerically, with the frequency-dependent masses
\beq
    X_0 &\equiv& \sqrt{\omega_q^\kappa\bar\omega_q^\kappa}, \phantom{|\bar\omega_q^\kappa|} X \equiv \sqrt{(\omega_q^\kappa)^2+m^2},\\
    \bar X_0 &\equiv& |\bar\omega_q^\kappa|, \phantom{\sqrt{\omega_q^\kappa\bar\omega_q^\kappa}}  \bar X \equiv \sqrt{(\bar{\omega}_q^\kappa)^2+m^2}, \\
    X_\pm &\equiv& \sqrt{X_0^2+\frac{m^2}{2}\Bigg(1\pm\sqrt{1+4\frac{X_0^2-\bar X_0^2}{m^2}}\Bigg)}\,,
\eeq
and where we have introduced the shifted, bosonic Matsubara frequencies $\smash{\omega_q^\kappa\equiv (2\pi q+\kappa\cdot r)T}$ and $\bar\omega_q^\kappa\equiv (2\pi q+\kappa\cdot \bar r)T$.

As for the matter contribution, at this order, it does not depend on $\bar r$ and corresponds to the well-known formula of the quark potential in the presence of a background $r$:
\beq
V^{\rm matter}(r) & = & -\frac{T}{\pi^2}\sum_{f,\rho}\int_0^\infty dq\,q^2\nonumber\\
& & \times\, \bigg[\ln\Big(1+e^{-\beta(\varepsilon_q^f+\mu+ir\cdot\rho T)}\Big)\nonumber\\
& & \hspace{0.5cm}+\,\ln\Big(1+e^{-\beta(\varepsilon_q^f-\mu-ir\cdot\rho T)}\Big)\bigg]\,,\label{eq: Vmatt}
\eeq
with $\smash{\varepsilon_q^f=\sqrt{q^2+M_f^2}}$. The summation symbol $\sum_{f,\rho}$ represents a sum over the fundamental (or defining) weights $\rho$ and the quark flavors $f$.

In the first line of  Eq.~\eqref{eq: Vglu}, $Z_aZ_{m^2}$ is a product of renormalization factors that will cancel the UV divergences but can contain different finite parts depending on the employed renormalization scheme. As mentioned before, we shall consider two popular schemes within CF model calculations: the vanishing momentum (VM) scheme and the infrared (IR) safe scheme \cite{Tissier2011InfraredCorrelators}. 

Each scheme is characterized by a running of the parameters $m$ and $g$ with the renormalization scale, which we denote $s$ in what follows, not to confuse it with the chemical potential $\mu$. This running is determined by integrating the corresponding beta functions from some initial conditions $m_0$ and $g_0$ (at a reference scale $s_0$), which are fitted to reproduce the lattice, zero-temperature ghost and gluon propagators. 

It can be argued that, in both schemes, the quarks do not contribute to $Z_a Z_{m^2}$ at this order. However, they enter the expression for $Z_a$ and the corresponding anomalous dimension. This affects the running of the parameters $m$ and $g$ and their initialization. We note however that the quark corrections to the anomalous dimensions are suppressed by a factor $s/M_f^2$ at large $M_f^2$, and thus, as a first approximation, we will ignore the effect of the quark on the parameter determination and running, and use those parameters obtained from fits of the pure Yang-Mills propagators.\footnote{Of course, this has some impact on the quantitative comparison with the simulations.} For the various schemes considered in this work, the initial parameters at $s_0=1\,$GeV are
\beq
    \text{IR safe}:&\ \quad &m_0=390\,\text{MeV},\quad g_0=3.7,\\
    \text{VM }\alpha=1:&\ \quad &m_0=500\,\text{MeV},\quad g_0=4.3,\\
    \text{VM }\alpha=2:&\ \quad &m_0=500\,\text{MeV},\quad g_0=4.7. \label{eq: schemes parameters}
\eeq
A more complete analysis would involve fitting propagators for distinct values of the quark masses as one explores the top-right corner of the Columbia plot. Unfortunately, there is no lattice data for the two-point functions spanning this whole region of the Columbia plot.

Once the parameters are fixed, the actual values of $r_3$ and $r_8$ will be obtained from the extremization of $V_{\bar r}(r)$ with respect to $r$ at fixed $\bar r$. The choice of the relevant physical extremum will depend on the nature of the chemical potential, see below. In what follows, we shall restrict the background to $\smash{\bar r=\bar r_c=(4\pi/3,0)}$, and so extremize $\smash{V_c(r)\equiv V_{\bar r_c}(r)}$. This choice originates in that the gluonic part of $V_c(r)$ is manifestly center-symmetric, and the relevant extremum $r$ becomes an order parameter for center-symmetry breaking in the limit of infinitely heavy quarks. A different, popular strategy is to extremize instead the background potential
\beq
\tilde V(\bar r)\equiv V_{\bar r}(r=\bar r)\,,\label{eq:Vtilde}
\eeq
whose gluonic part, as given by the first line of Eq.~\eqref{eq: Vglu}, is also manifestly center-symmetric. Although this function does not correspond to a Legendre transform of a thermodynamical potential, it can be argued that its extremization also gives access to the gauge-field average, in a particular gauge where this gauge-field average is always equal to the background. However, the proof of this statement involves additional identities compared to the case of a standard Legendre transform. As these identities can be partially broken in the presence of approximations or modeling, we expect this approach to yield less sensible results, see also \cite{MariSurkau2024DeconfinementDependences}. We shall compare the two approaches below.

In addition to extracting the values of $r_3$ and $r_8$, we will evaluate the corresponding Polyakov loops $\ell$ and $\bar\ell$. For simplicity, we shall consider the tree-level relation between the two, which is found to be 
\beq
\ell & = & \frac{1}{3}\Bigg[e^{-i\frac{r_8}{\sqrt{3}}}+2e^{i\frac{r_8}{2\sqrt{3}}}\cos(r_3/2)\Big]\,,\label{eq: l of r}\\
\bar\ell & = & \frac{1}{3}\Bigg[e^{i\frac{r_8}{\sqrt{3}}}+2e^{-i\frac{r_8}{2\sqrt{3}}}\cos(r_3/2)\Big]\,.\label{eq: lb of r}
\eeq
We stress that these expressions are valid no matter what Weyl chamber $\smash{r=(r_3,r_8)}$ lies in. In practice, however, the relevant extremum value of $r$ will be found in the same Weyl chamber as $\bar r$. In this case, there is a one-to-one relation between the values of $\ell$ and $\bar\ell$ and those of $r_3$ and $r_8$.\footnote{In the case of non-vanishing real chemical potential, $r_8$ becomes imaginary, see below, and there is no notion of Weyl chambers anymore.} Also, the symmetry constraints derived above for the Polyakov loops translate into symmetry constraints for $r_3$ and $r_8$, see the Appendix for details. These constraints are useful as they indicate where to look for the relevant extrema of the potential.

%%%%%
\section{Results\label{sec: results}}
Let us now collect and discuss our results for the phase structure in the heavy-quark regime.

\begin{figure}[t]
\includegraphics[height=0.24\textheight]{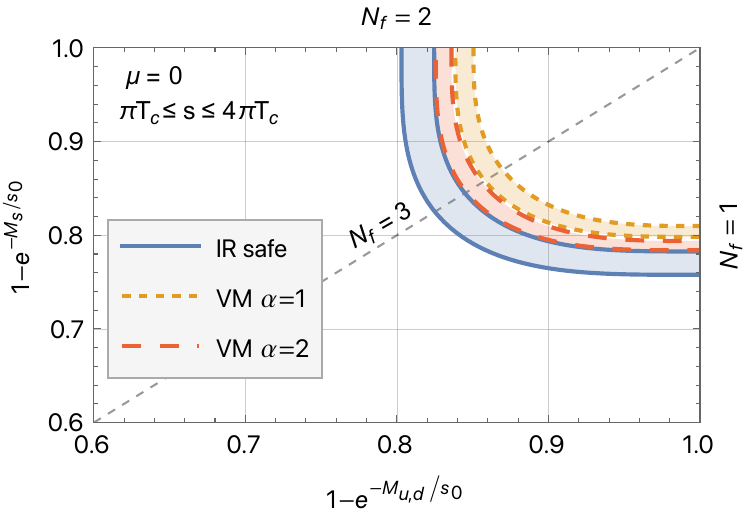}
\caption{Critical boundary in the heavy-quark corner of the Columbia plot as determined from the center-symmetric potential $V_c(r)$. The normalization $s_0$ refers to the scale at which the renormalization group flow is initialized, here $\smash{s_0=1}\,$GeV. For each of the considered renormalization schemes, the critical boundary moves in the thin bands as the renormalization scale is varied from $\pi T_c$ to $4\pi T_c$.}
\label{fig: Columbia 0mu Vc}
\end{figure}

%%%
\subsection{Zero chemical potential\label{subsec: 0 mu results}}
In the case of a vanishing chemical potential, the relevant extremum of $V_c(r)$ is its global minimum, and we can assume that $\smash{r^8=0}$ from charge-conjugation invariance, see the Appendix. 

For infinitely large quark masses, the potential boils down to the glue potential whose minimum experiences a first-order, discontinuous transition as a function of temperature, located between two spinodal temperatures at which the curvature of $V_c(r)$ vanishes for some $r$. As the quark masses are lowered, the matter contribution to the potential kicks in, and the two spinodal temperatures get closer and closer to each other until they eventually become equal. For any such values of the quark masses, the minimum of the potential experiences a second-order, continuous transition. This defines a critical boundary in the upper right corner of the Columbia plot, which separates a first-order region containing the YM limit from a crossover region containing the QCD physical point.

To determine the critical boundary, we have to find, as a function of the $u$- and $d$-quark masses (assumed to be equal for simplicity), the values of the temperature $T_c$, the $s$-quark mass $M_s$, and the restricted gauge field average $r^3_c$ where the lower and upper spinodals coincide. This happens when 
\begin{equation}\label{eq: 0mu eqns}
    V_c^\prime(r)=0, \quad V_c^{\prime\prime}(r)=0, \quad V_c^{\prime\prime\prime}(r)=0\,,
\end{equation}
where the primes denote successive derivatives with respect to $r_3$-derivatives while $r_8$ is implicitly taken equal to $0$.

\begin{figure}[t]
\includegraphics[height=0.22\textheight]{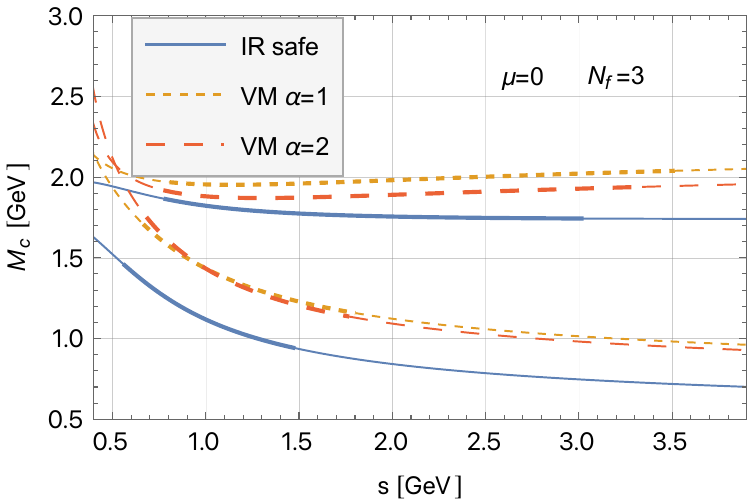}
\caption{Renormalization scale dependence of the critical quark mass in the case of three degenerate quark flavors at vanishing chemical. The three upper, flatter lines were evaluated using the center-symmetric potential $V_c$, the lower ones using the background effective potential $\Tilde{V}(\bar{r})$. The thicker lines highlight the regions where $\pi T_c\leq s\leq4\pi T_c$.}
\label{fig: Mc 0mu}
\end{figure}

In Fig.~\ref{fig: Columbia 0mu Vc}, we show the results for the heavy-quark critical boundary found using the center symmetric potential $V_c(r)$. We use some appropriate variables that allow one to bring the infinite quark mass limit into a finite range. For each of the renormalization schemes introduced in the previous section, we show the range of positions of the critical boundary as the renormalization scale $s$ is varied in the range $\pi T_c\leq s\leq 4\pi T_c$, a conventional range in finite temperature calculations. Note that the critical temperature is found to be pretty much insensitive to the location along the boundary line or to the chosen scheme. The first feature is expected in the heavy-quark region, see for instance Ref.~\cite{Maelger2018UniversalQuarks}, while the second is a welcome feature of the center-symmetric CF model, see Ref.~\cite{MariSurkau2024DeconfinementDependences}, which signals that the present one-loop approximation is under control. 

We find that the critical boundary has only a mild dependence on the chosen renormalization scheme or the scale $s$ over the considered range (and even beyond). This can also be seen in Fig.~\ref{fig: Mc 0mu} where we show the renormalization scale dependence of the critical mass for three degenerate quark flavors. Since $T_c\simeq 260\,$MeV, the range over which $s$ is varied is approximately $s\in[0.8,3.3]\,$GeV with a slight variation in the endpoint between $3-3.5\,$GeV for the different schemes. That the critical masses turn out to be almost scheme-independent is another welcome feature, in line with the fact that they correspond to bare quark masses at this level of approximation, see further remarks below.

We have also checked that, to a high degree of accuracy, the critical boundary satisfies the relation 
\begin{equation}\label{eq: Columbia line shape}
    2f(M_{u,d}/T_c)+f(M_s/T_c)=3f(M_3/T_c)\,,
\end{equation}
where $M_3$ is the critical mass in the case of three degenerate quark flavors, and $f(x)=(3x^2/\pi^2)K_2(x)$, with $K_2(x)$ the modified Bessel function of the second kind. This result is expected in any approach that includes a one-loop matter contribution potential added to a phenomenologically modeled glue contribution \cite{Maelger2018UniversalQuarks}.

\begin{figure}[t]
\includegraphics[height=0.24\textheight]{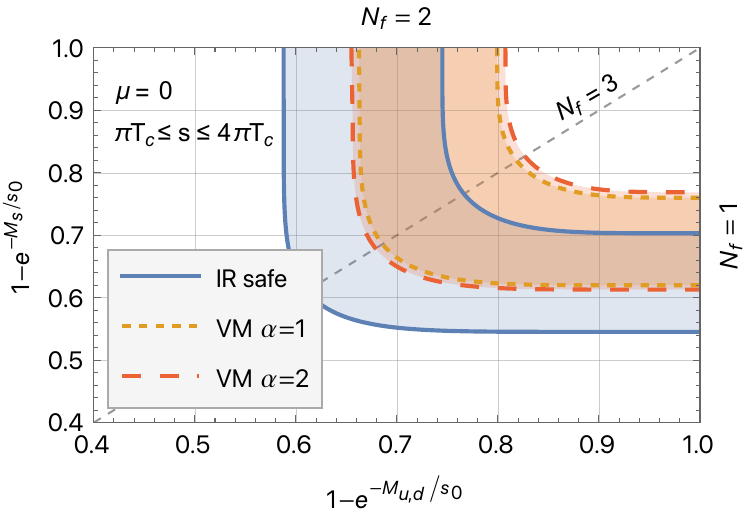}
\caption{Same as in Fig.~\ref{fig: Columbia 0mu Vc} using the background effective potential $\Tilde{V}(\bar{r})$.}
\label{fig: Columbia 0mu Vb}
\end{figure}

In Fig.~\ref{fig: Mc 0mu}, we also show the critical mass as obtained from the background effective potential $\Tilde{V}(\bar{r})$, taking once more the case of three degenerate quark flavors for illustration. The corresponding Columbia plot is shown in Fig.~\ref{fig: Columbia 0mu Vb}. The critical boundary in this case is farther away from the top right corner (note the different scale compared to Fig.~\ref{fig: Columbia 0mu Vc}). This is because the critical masses found using $\Tilde{V}(\bar{r})$ are lower, see Fig.~\ref{fig: Mc 0mu}. All the curves follow the same universal relation \eqref{eq: Columbia line shape}, but the dependence on the renormalization scale is much larger than with $V_c(r)$ as illustrated by the much thicker bands. While, as expected, there is a spurious dependence on the renormalization scale and the renormalization scheme in both cases, for the center-symmetric potential the critical masses only vary by around $5\%$ in each scheme and are also closer between the IR safe scheme and VM schemes. For the background effective potential, the variation is instead of around $50\%$, and the IR safe scheme is further separated from the VM scheme. This confirms the expectation that the one-loop approximation in the case of the center-symmetric potential is better controlled than in the case of the background effective potential. 

The fact that, at this order, the extracted quark masses can be interpreted as bare quark masses allows for a direct comparison with the numerical simulations. Our predictions for the ratio $R_3=M_3/T_c$ are collected in Table \ref{tab: R_3 mu0}. They lie roughly $10\%$ away from the lattice simulations, which is a satisfactory result given the one-loop nature of the approximation and the fact that our parameter determination is based on pure YM results. A further improvement would include computing two-loop corrections but one should mention that, in this case, the comparison to the simulations cannot be made directly in terms of the $R_{N_f}$'s but rather, in terms of ratios of $R_{N_f}$'s for different $N_f$'s \cite{Maelger2018PerturbativeCorrections}.

\begin{table}[h]
    \centering
    \begin{tabular}{c|c|c|c|c|c}
             &\ IR safe\ \ & VM $\alpha=1$ & VM $\alpha=2$ &\ $\Tilde{V}(\bar{r})$\ \ &\ Lattice\ \ \\ \hline
         $R_3$ & $7.31\pm0.05$ & $7.37\pm0.09$ & $7.40\pm0.11$ & 8.07 & 8.32
    \end{tabular}
    \caption{Values of $\smash{R_3\equiv M_3/T_c}$ at $\smash{\mu=0}$ predicted from $V_c(r)$ as obtained within the three considered schemes with their standard deviation across $s\in[\pi T_c,4\pi T_c]$. $R_3$ is monotonically decreasing as $s$ increases in this interval in all cases. For comparison, we also give the values obtained from $\Tilde{V}(\bar{r})$ and from the simulations.}
    \label{tab: R_3 mu0}
\end{table}

Finally, we mention that the prediction for the ratio $R_3$ coming from $\tilde V(\bar r)$ seems slightly better, even though the critical masses are smaller than those predicted from $V_c(r)$. This hides the fact that the transition temperature is strongly underestimated in the case of $\tilde V(\bar r)$, where it is of the order of $100$ MeV. The actual temperature in this range of the Columbia plot should be close to the transition temperature in pure YM theory, of the order of $270\,$MeV, as it is when using $V_c(r)$. 

Also, the ratio $R_3$ obtained from $\tilde V(\bar r)$ at one-loop is identically scale and scheme-independent. This is due to the relative simplicity of $\Tilde{V}(\bar{r})$ at this order. Indeed, since there is no explicit coupling dependence, $M_c$ follows the running of $m$ up to an overall factor. The same is true for the critical temperature $T_c$, resulting in a constant ratio $R_3$. Because this will not be true anymore upon including higher order corrections, this adds to our suspicion that the one-loop approximation to $\tilde V(\bar r)$ is too simplistic at this order.

In what follows, we consider exclusively the center-symmetric potential $V_c(r)$. For more results using the background potential $\tilde V(\bar r)$, see for instance \cite{Reinosa2015PerturbativePotential}.
From now on, we shall also assume $N_f$ degenerate quarks. The generalization to the whole boundary in the Columbia plot can be made via Eq.~\eqref{eq: Columbia line shape}.

%%%
\subsection{Imaginary chemical potential\label{subsec: im mu results}}
With a non-zero chemical potential, whether real or imaginary, charge-conjugation symmetry is explicitly broken, and we can no longer assume $\smash{r^8=0}$. The imaginary chemical potential case $\mu=i\mu_i$ presents an interesting phase structure due to the symmetries discussed in Sec.~\ref{sec:syms}. The positivity of the determinant implies that the center-symmetric potential is real and admits a global minimum for a real $\smash{r^8\in\R}$. From Eqs.~\eqref{eq: l of r}-\eqref{eq: lb of r} we see that with real and non-zero $r^8$ the Polyakov loops themselves will be complex (conjugates), as required by the symmetry constraint \eqref{eq:lim}. Also, due to the various transformations discussed in Sec.~\ref{sec:transfos} that connect different chemical potential values, we can restrict our analysis to $\mu_i/T\in[0,\pi/3]$.

As we increase $\mu_i/T$, the critical boundary moves in the Columbia plot. Since $\smash{r_8}$ does not vanish anymore, we need to generalize the equations \eqref{eq: 0mu eqns} that determine this boundary line. The correct equations are
\begin{equation}\label{eq: finite mu eqns}
    \nabla V(r)=0,\quad |\mathcal{H}(V(r))|=0,\quad [v\cdot\nabla]^3V(r)=0\,,
\end{equation}
where $\smash{\nabla=(\partial_3,\partial_8)}$ ($\partial_j\equiv\partial/\partial r^j$) is the gradient in the $r$-plane, $|\mathcal{H}(V(r))|=(\partial_{33}V)(\partial_{88}V)-(\partial_{38}V)^2$ is the determinant of the Hessian of $V$ at $r$, and $v$ is the eigenvector of the Hessian with zero eigenvalue. Explicitly, the last equation reads
\beq
    0=&&(\partial_{88}V)^3 (\partial_{333}V) 
    -3(\partial_{88}V)^2 (\partial_{38}V)(\partial_{338}V)\nonumber\\
    &&+3(\partial_{88}V)(\partial_{38}V)^2 (\partial_{388}V)
    -(\partial_{38}V)^3 (\partial_{888}V)\,.
\eeq
In principle, these equations serve the same purpose as \eqref{eq: 0mu eqns}. The vanishing gradient ensures an extremum in the $r$-plane, and the vanishing of the other two quantities ensures it is both a saddle point and a critical point, which only occurs for the critical parameters for which the two spinodals combine into a minimum, signaling the change from first-order transition to crossover.

As $\mu_i/T$ approaches $\pi/3$, the critical boundary moves towards the Roberge-Weiss boundary, characterized by tricritical behavior. This has been well documented in the literature, and the $\mu$-dependence of the critical mass and temperature ratio $R_{N_f}$ in the vicinity of the tricritical point is described by the scaling law
\begin{equation}\label{eq: crit scaling}
    R_{N_f}(\mu)=\frac{M_{\rm tri}}{T_{\rm tri}} + K x^{2/5},
\end{equation}
where
\begin{equation}
    x = \left(\frac{\pi}{3}\right)^2-\left(\frac{\mu_i}{T}\right)^2 = \left(\frac{\pi}{3}\right)^2+\left(\frac{\mu}{T}\right)^2,
\end{equation}
and $M_{\rm tri}$ and $T_{\rm tri}$ are the critical parameters at $\mu_i/T=\pi/3$. By rewriting $x$ in terms of a general chemical potential $\mu$ instead of $\mu_i$ the scaling relation can be extended to give predictions for the critical parameters at arbitrary real chemical potentials $\mu\in\R$, see below.

\begin{figure}[t]
    \centering
    \includegraphics[width=0.95\linewidth]{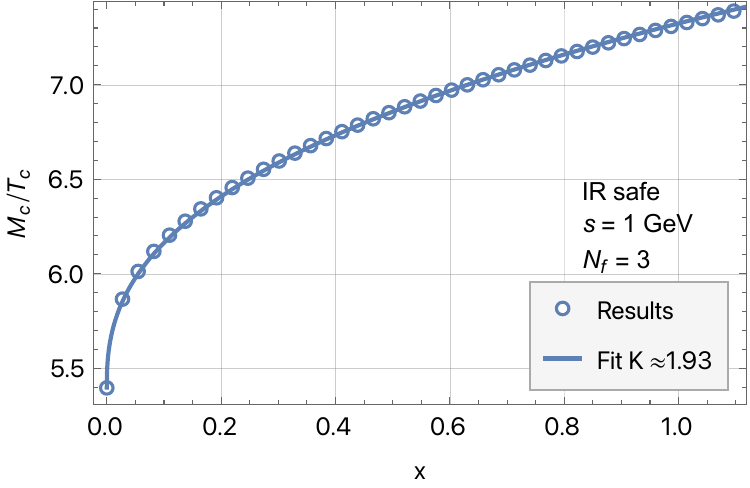}
    \caption{A fit of our results for $\smash{R_3=M_c/T_c}$ by the scaling law \eqref{eq: crit scaling}, with $x=\left(\pi/3\right)^2-\left(\mu_i/T\right)^2$. We used the IR safe scheme at the scale $s=1\,$GeV.}
    \label{fig: R3 mu-im}
\end{figure}

The tricritical point can be found by following the trajectory from $\mu=0$ to $\mu=i\pi T/3$ while solving Eqs.~\eqref{eq: finite mu eqns}. However, due to the singular scaling of $x^{2/5}$ as $x\to0$, it is more convenient to work directly at $\mu=i\pi T/3$ and to adapt the equations appropriately. Even better, as we mentioned above, without loss of generality, the Roberge-Weiss transition can be studied at $\mu=i\pi T$. In this case, the Roberge-Weiss symmetry means that the potential is invariant under $r_8\to -r_8$, and an explicitly realized symmetry would impose $\smash{r_8=0}$ while the breaking occurs in the $r_8$ direction.

Introducing the function $r_3(r_8)$ such that
\beq
0=\partial_3 V|_{r_3=r_3(r_8)}\label{eq:eom}
\eeq
and the reduced potential
\beq
\hat V(r_8)\equiv V(r_3(r_8),r_8)\,,
\eeq
the conditions defining the boundary are now \eqref{eq:eom} together with
\beq
0=d_{88}\hat V|_{r_8=0}=d_{8888}\hat V|_{r_8=0}\,.
\eeq
Note that the cubic derivative does not appear because of the $r_8\to -r_8$ symmetry. These three conditions can be rewritten in terms of $V(r)$ as
\begin{equation}
    0=\partial_3V = \partial_{88}V = -3(\partial_{388}V)^2+(\partial_{33}V)(\partial_{8888}V)\,,
\end{equation}
where it is understood that the derivatives are evaluated for $\smash{r_8=0}$.

\begin{table}[h]
    \centering
    \begin{tabular}{c|c|c|c|c|c}
             &\ IR safe\ \ & VM $\alpha=1$ & VM $\alpha=2$ &\ $\Tilde{V}(\bar{r})$\ \ &\ Lattice\ \ \\ \hline
         $R_3$ & $5.32\pm0.05$ & $5.38\pm0.10$ & $5.39\pm0.11$ & 6.14 & 6.66
    \end{tabular}
    \caption{Values of $\smash{R_3\equiv M_3/T_c}$ at $\smash{\mu=i\pi T/3}$ predicted from $V_c(r)$ as obtained within the three considered schemes with their standard deviation across $s\in[\pi T_c,4\pi T_c]$. $R_3$ is monotonically decreasing as $s$ increases in this interval in all cases. For comparison, we also give the values obtained from $\Tilde{V}(\bar{r})$ and from the simulations.}
    \label{tab: R_3 mu_i}
\end{table}

In Fig.~\ref{fig: R3 mu-im}, we show an example of how the critical ratio $R_3=M_c/T_c$ varies as a function of $x=\left(\pi/3\right)^2-\left(\mu_i/T\right)^2$, along with the fit according to \eqref{eq: crit scaling}. We have checked that the quality of the fit is the same as the one shown here for the other renormalization schemes, a wide range of renormalization scales, and other numbers of degenerate quarks $N_f$. Note that the chemical potentials at which we solved the equations are evenly spaced in $x$. However, the scaling results in rapid variations of $R_3$ towards the tricritical point. Hence, it is more efficient to solve directly for the tricritical point than trying to approach it.

\begin{figure}[t]
    \centering
    \includegraphics[width=0.95\linewidth]{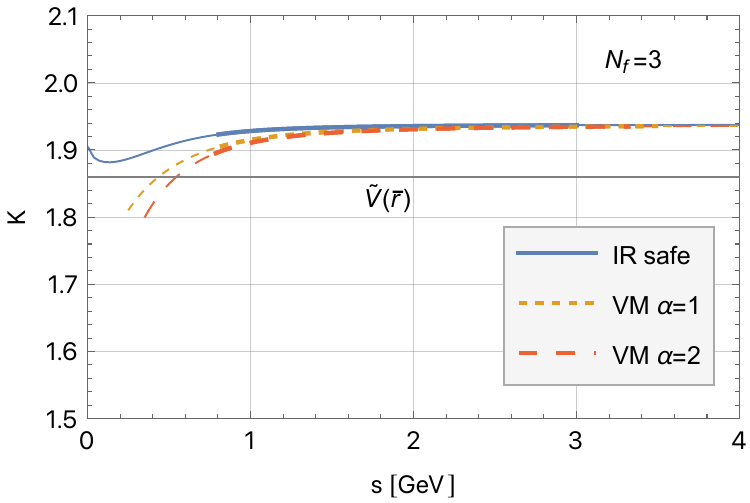}
    \caption{Fit parameter $K$ as a function of renormalization scale $s$ in the different schemes. The lines are slightly thicker in the regions where $\pi T_c\leq s\leq4\pi T_c$.}
    \label{fig: K of s}
\end{figure}

In Fig.~\ref{fig: K of s}, we show the renormalization scale dependence of the fitted scaling parameter $K$, see Eq.~\eqref{eq: crit scaling}, for three degenerate quarks. We show the results in the three different considered schemes and highlight the range of scales $\pi T_c\leq s\leq4\pi T_c$ by slightly thicker lines. Note that the results for $K$ depend even less on the renormalization scale and renormalization scheme than the critical temperature or the critical quark masses. The coefficient is essentially constant, $\smash{K\approx1.93}$. The slightly larger variations happen for small scales, where special care has to be taken anyway because the coupling in the VM schemes becomes non-perturbative around $\smash{s=0.5-0.6\,}$GeV as it approaches the Landau pole. The IR-safe coupling, while not presenting any Landau pole, also grows to non-perturbative values at $\smash{s=0.2\,}$GeV before decreasing again towards zero in the deep IR.

The lattice value for $K$ at $\smash{N_f=1}$ is 1.55 \cite{Fromm2012ThePotentials}. To our knowledge, no lattice data is available at $\smash{N_f=3}$. We have also calculated $K$ for $\smash{N_f=2}$ and $\smash{N_f=1}$, giving roughly 1.97 and 2.05, with the same shape, roughly constant, as shown in Fig~\ref{fig: K of s}.\\

\begin{figure}[t]
    \centering
    \includegraphics[width=0.8\linewidth]{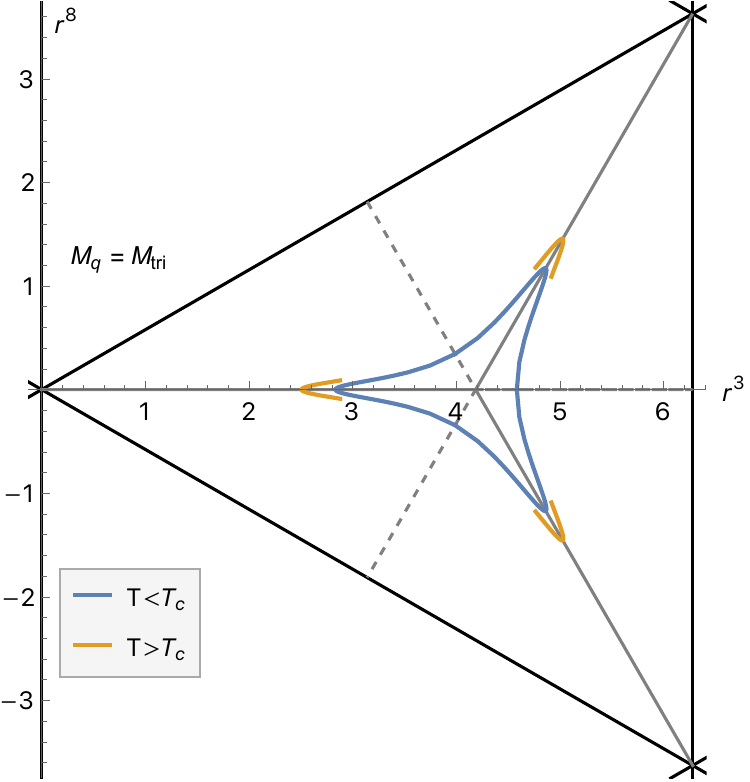}
    \caption{Trajectories of the global minimum $r_{\rm min}(\mu_i)$ as $\mu_i$ varies from 0 to $2\pi T$, at fixed temperatures $T$ above and below the Roberge-Weiss transition temperature $T_c$. They move clockwise as $\mu_i$ increases.}
    \label{fig: mu_i min trajectories}
\end{figure}

Finally, it can be useful to visualize the Roberge-Weiss transition in the Weyl chamber. In Fig.~\ref{fig: mu_i min trajectories}, we show the trajectory followed by the global minimum $r_{\rm min}(\mu_i)$ of $V_c(r)$ as the imaginary chemical potential $\mu_i$ varies from\footnote{It is enough to obtain the trajectory for $\mu_i/T$ varying from $0$ to $\pi/3$ and to deduce the rest using the various symmetries discussed in the Appendix.} 0 to $2\pi T$ in the case where the quark mass is adjusted to its tricritical value $M=M_{\rm tri}$ (or below).\footnote{For $M_q>M_{\rm tri}$ and $T_c(\mu=0)<T<T_c(\mu_i=\pi T/3)$, the trajectories are more complicated but can be analyzed similarly.} 

If $\smash{T<T_c}$, the Roberge-Weiss symmetry is manifest and the minimum moves continuously clockwise around the center-symmetric point $\smash{r=\bar r_c=(4\pi/3,0)}$.\footnote{To create Fig.~\ref{fig: mu_i min trajectories} the $T<T_c$ was chosen practically equal to $T_c$. As $T$ decreases, the trajectory quickly shrinks towards the center-symmetric point $\bar{r}_c$. The same is true if we increase $M_q$ instead.} Introducing $\smash{\Delta r\equiv r-\bar r_c}$ and then $\smash{\Delta z\equiv\Delta r_3+i\Delta r_8}$, we find that the phase of $\Delta z$ is $(\pi/3)k$ when $\mu_i/T=(\pi/3)k$, in agreement with the constraints from ${\cal KZ}$-symmetry, see the Appendix. The symmetry only fixes the phase modulo $\pi$. Here, we find that the phase seems to be fixed in one specific way. In particular, while $\smash{r_8=0}$ for both $\smash{\mu_i/T=0}$ and $\smash{\mu_i/T=\pi}$, we find $\smash{0<r_3<4\pi/3}$ in the first case and $\smash{4\pi/3<r_3<2\pi}$ in the second. In terms of the Polyakov loop, this corresponds to $\smash{\ell=\ell^*>0}$ and $\smash{\ell=\ell^*<0}$ respectively.

If $\smash{T>T_c}$, the Roberge-Weiss symmetry is spontaneously broken at the chemical potentials $\mu_i/T=(pi/3)k$. In this case, the trajectory of the global minimum becomes discontinuous and avoids the points $\Delta r$ whose phase $(\pi/3)k$ with odd $k$. In particular, $\ell$ is not anymore real for $\smash{\mu_i/T=\pi}$, while it remains real and positive for $\smash{\mu_i/T=0}$.

%%%
\subsection{Real chemical potential\label{subsec: real mu results}}
One extra difficulty occurs in the case of a real chemical potential. Indeed, because the fermionic determinant becomes complex, thermal averages of real-valued quantities do not need to be real-valued. In particular, this is the case for the expectation values $r_3$ and $r_8$.

We show in the Appendix that, for the choice of background $\smash{\bar r=\bar r_c=(4\pi/3,0)}$, $r_3$ is real while $r_8$ is imaginary.\footnote{Similarly, it was argued in Ref.~\cite{Reinosa2015PerturbativePotential}, that, in the background effective potential approach $\bar r_3$ needs to be taken real while $\bar r_8$ needs to be taken imaginary.} Using Eqs.~\eqref{eq: l of r} and \eqref{eq: lb of r}, this is checked to be consistent with $\ell$ and $\bar\ell$ being both real (and independent of each other). For those values of $r_3$ and $r_8$, it can also be argued, see the Appendix, that the potential remains real, which is a welcome feature if the relevant extremum is to be related to thermodynamical observables. However, this extremum has no reason to be a minimum; in fact, it is generally a saddle point. As long as this saddle point is the only extremum, there is no ambiguity in the choice of extremum, but whenever two saddles coexist, it becomes unclear which one should be chosen. The non-positivity of the fermionic determinant makes it difficult to identify a general strategy to select the relevant extremum.

In analogy to the cases with multiple minima, we choose the deepest saddle point, i.e., the one with the smaller $V(r)$ value. This recipe is certainly the correct one at $\mu=0$ since in this case one can choose to work either with $V(r^3,r^8,\mu=0)$ or with $V(r^3,ir^8_i,\mu=0)$. The absolute minimum of the first potential, which occurs for $r_8=0$, appears as the deepest saddle point for the second potential. Keeping the same recipe for small, non-zero chemical potentials is certainly still the right choice, leveraging continuity. However, there is no guarantee that this always corresponds to the physically correct procedure for larger chemical potentials. This ambiguity could be qualified as a weak sign problem in the continuum.

Following the above recipe, the fit \eqref{eq: crit scaling} of $R_{N_f}(\mu)$ for imaginary chemical potentials can be extrapolated to the real chemical potential range and compared to the result of solving Eqs.~\eqref{eq: finite mu eqns} in that same range. The additional factors of $i$ from the $r^8_i$ derivatives cancel out, resulting in the same equations.
\begin{figure}[t]
    \centering
    \includegraphics[width=0.95\linewidth]{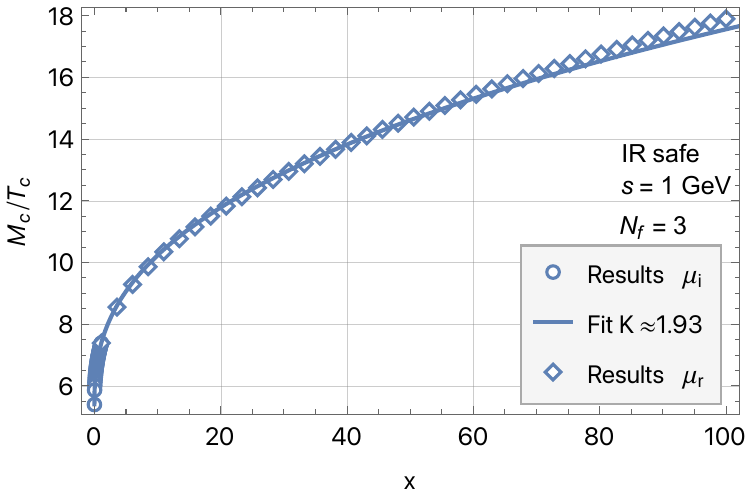}
    \caption{Critical ratio $R_3(\mu)=M_c(\mu)/T_c(\mu)$ as a function of $x$ showing both the imaginary $\mu$ regime (circles) and the real $\mu$ regime (diamonds). The points are the results found directly from the potential $V(r)$ while the line is the extrapolation from the imaginary $\mu$ fit shown above.}
    \label{fig: R3 mu-re}
\end{figure}
In Fig.~\ref{fig: R3 mu-re}, we show the results for the critical ratio $R_3(\mu)$ as a function of $x=(\pi/3)^2+(\mu/T)^2$ for both imaginary $0<x<(\pi/3)^2$ and real chemical potentials up to roughly $\mu\approx2.5\,$GeV. We also show the extrapolation of the fit performed only on the imaginary $\mu$ results. For consistency with Fig.~\ref{fig: R3 mu-im}, we chose the same renormalization scale ($s=1\,$GeV) and the same renormalization (IR safe) scheme, but, as we have checked, the plot looks almost identical for other choices as well.

We observe that the extrapolation of the critical scaling provides an accurate estimate of the critical parameters for real chemical potential. Towards larger $\mu$, the extrapolation starts underestimating the results, but the extrapolation is valid for a fairly large range of chemical potentials. In comparison to previous results obtained with the background effective potential $\Tilde{V}(\bar{r})$, see Ref.~\cite{Reinosa2015PerturbativePotential, Maelger2018PerturbativeCorrections}, we also note an improvement in the accuracy of the extrapolation when using $V(r)$, especially at larger $\mu$ it is slightly closer to the directly calculated results.

\section{Conclusions\label{sec: conclusions}}
In this work, we have tested the recently proposed center-symmetric Curci-Ferrari model-based approach to QCD at finite temperature and density beyond the pure Yang-Mills studies of Ref.~\cite{vanEgmond2022ATransition, MariSurkau2024DeconfinementDependences}. To this purpose, we included 2+1 flavors, assumed to be very heavy, and studied the corresponding critical boundary in the top right corner of the Columbia plot where the first-order confinement/deconfinement transition turns into a smooth crossover. This work also complements that of Ref.~\cite{Reinosa:2015oua} through the replacement of the background effective potential used in that reference by the center-symmetric effective potential of Ref.~\cite{vanEgmond2022ATransition, MariSurkau2024DeconfinementDependences}.

We have shown that, at one-loop order, our approach predicts already all qualitative features expected of the heavy-quark region, for imaginary and real chemical potentials, including the known scaling relations for the critical parameters. In the case of an imaginary chemical potential, we have emphasized the distinction between the Roberge-Weiss symmetry and the center symmetry and argued that the former could be present in systems without center symmetry. 

We have tested the quality of our one-loop approximation-based results by monitoring their renormalization scheme and scale dependence. We find only very small variations, at the $1\%$ level or lower, across a typical range of scales for the critical $M_c$ or critical temperature $T_c$ along the boundary line, which represents a clear improvement compared to previous approaches using the background effective potential.

While the quantitative results do not match results from lattice QCD exactly, they are not too far from them. It should also be kept in mind that our determination of the parameters is not ideal, as we do not have access to correlation functions across the whole heavy-quark region of the Columbia plot. An advantage of the presented model over lattice QCD is that it allows direct access to the finite chemical potential regime. We have used this to verify that the common extrapolation from imaginary chemical potentials is accurate up to large chemical potentials.

Of course, the pressing question is how well this approach operates in the physical QCD case, including light quarks. In this respect, one possible and easy extension of the present work is to couple the center-symmetric potential, a proxy for the Polyakov loop potential, to existing low-energy models that capture spontaneous chiral symmetry-breaking at a low cost. Many similar works exist, but they usually depend on a multi-parameter Polyakov loop potential. The benefit of our approach is that the potential depends on only one phenomenological parameter. The outcome of this analysis is work in progress and will be presented elsewhere. Another possibility offered by the CF model is to include the matter sector and the quark-gluon coupling in a first-principle fashion by taking advantage of a combined expansion in the coupling in the pure gauge sector and the inverse of the number of colors. Work in this direction is also in progress.

%%%%%
\appendix

%%%%%
\section{More on \texorpdfstring{$V_{\bar r}(r)$}{V(r)} and its symmetries\label{sec: Weyl chambers}}

It is first important to realize that, unlike the Polyakov loops $\ell$ and $\bar\ell$, the variables $r_3$ and $r_8$ that enter the potential $V_{\bar r}(r)$ are not observables. This means that the value of $\smash{r=(r_3,r_8)}$ can be changed by using gauge transformations. Of course, by construction, gauge transformations are not symmetries of the gauge-fixed action and what happens, in general, is that the value of the background $\smash{\bar r=(\bar r_3,\bar r_8)}$ is changed as well, what can be interpreted as the fact that gauge transformations connect different gauges. 

In the present case, the relevant gauge transformations are all generated by the particular subclass
\beq
r & \to & 4\pi\alpha\,p+r-2\frac{r\cdot\alpha}{\alpha^2}\alpha\equiv r_{\alpha,p}\,,\label{eq:gg1}\\
\bar r & \to & 4\pi\alpha\,p+\bar r-2\frac{\bar r\cdot\alpha}{\alpha^2}\alpha\equiv \bar r_{\alpha,p}\,,\label{eq:gg2}
\eeq
where $\alpha$ denotes any root of SU(3), that is any non-zero adjoint weight, and $\smash{p\in\mathds{Z}}$. Geometrically, these transformations correspond to reflection symmetries with respect to hyperplanes orthogonal to $\alpha$ and displaced from the origin by $2\pi\alpha\,p$.\footnote{In the case of SU(3), these hyperplanes are just straight lines but in the SU(N) case they would correspond to $(N-2)$-dimensional subspaces in the space of $r$ or $\bar r$ which is of dimension $N-1$.} The corresponding network of hyperplanes subdivides each of the spaces of the variables $r$ and $\bar r$ into the Weyl chambers referred to in the main text and the invariance property
\beq
V_{\bar r}(r)=V_{\bar r_{\alpha,p}}(r_{\alpha,p})\label{eq:red}
\eeq
just reflects the fact that the Weyl chambers are all physically equivalent.

Let us now reconsider the various symmetries discussed in Secs.~\ref{sec:transfos}-\ref{sec:rmqs} from the point of view of $V_{\bar r}(r)$. This will allow us to determine constraints on the relevant extremum $r$. As in Sec.~\ref{sec:syms}, we assume that no symmetry is spontaneously broken, as in the interior of the Columbia plot, because both center and chiral symmetry are explicitly broken. In this case, the symmetries apply to all the extrema, even in the presence of spinodal branches.

It is also important to realize that, because the transformations \eqref{eq:gg1}-\eqref{eq:gg2} correspond to redundancies of the description of the system in terms of gauge fields, any physical symmetry can be redefined, upon convenience, modulo any combination of these redundant transformations.

Let us consider, for instance, a ${\cal C}$-transformation. In its usual form, it corresponds to $(r_3,r_8)\to(-r_3,-r_8)$ and $(\bar r_3,\bar r_8)\to(-\bar r_3,-\bar r_8)$ but, upon combining it with one of the redundant transformations, it can be recast into $(r_3,r_8)\to(r_3,-r_8)$ and $(\bar r_3,\bar r_8)\to(\bar r_3,-\bar r_8)$. Alongside the change $\smash{\mu\to -\mu}$ associated to ${\cal C}$, this leads to
\beq
V_{\bar r_3,\bar r_8}(r_3,r_8;\mu)=V_{\bar r_3,-\bar r_8}(r_3,-r_8;-\mu)\,.
\eeq
When choosing the background as $\smash{\bar r=\bar r_c=(4\pi/3,0)}$, which is invariant under the considered transformation, it follows that
\beq
V_c(r_3,r_8;\mu)=V_c(r_3,-r_8;-\mu)\,.
\eeq
The extremum then obeys the relation
\beq
r_3(-\mu)=r_3(\mu) \quad \mbox{\rm and} \quad r_8(-\mu)=-r_8(\mu)\,,\label{eq:rC}
\eeq
which should be put into correspondence with Eq.~\eqref{eq:lC}. In particular, for $\smash{\mu=0}$, this implies that
\beq
r_8(\mu=0)=0\,,
\eeq
which corresponds to Eq.~\eqref{eq:lC2}. 

The ${\cal A}$-transformations imply
\beq
V_{\bar r}(r;\mu)=V_{\bar r}(r;\mu+i2\pi p T)\,.
\eeq
and thus
\beq
r(\mu+i2\pi pT)=r(\mu)\,.
\eeq
Combining this with Eq.~\eqref{eq:rC} above, one arrives at
\beq
r_3(2i\pi pT-\mu) & = & r_3(\mu)\,,\\
r_8(2i\pi pT-\mu) & = & -r_8(\mu)\,,
\eeq
and thus, for $\mu=i\pi p T$, we find the constraint
\beq
r_8(i\pi pT)=0\,.\label{eq:rCA}
\eeq
As before, the status of this identity depends on the parity of $p$. For $p$ even, the identity is always fulfilled because the symmetry constraint at $\smash{p=0}$ stems directly from ${\cal C}$ which is not expected to break spontaneously, and the other even values of $p$ are connected to $p=0$ by the ${\cal A}$-transformations. In contrast, for odd $p$, the symmetry can break spontaneously because the identity for $p=1$ stems from ${\cal CA}$ and neither ${\cal C}$ nor ${\cal A}$ are symmetries of the action for the corresponding value of $\mu$. The other odd values of $p$ can again be reached from $p=1$ upon use of ${\cal A}$.

As for the ${\cal Z}$-transformations, up to an appropriate choice of redundant transformations \eqref{eq:gg1}-\eqref{eq:gg2} they can be written as
\beq
& & V_{\bar r}(r;\mu)\nonumber\\
& & \hspace{0.2cm}=\,V_{\bar r_c+R_k\cdot(\bar r-\bar r_c)}(\bar r_c+R_k\cdot (r-\bar r_c);\mu+i(2\pi/3) k T)\,,\nonumber\\
\eeq
where $R_k$ is the planar rotation by an angle $(2\pi/3)k$. Choosing $\bar r=\bar r_c$, this becomes
\beq
V_c(r;\mu)=V_c(\bar r_c+R_k\cdot (r-\bar r_c);\mu+i(2\pi/3) k T)\,.\nonumber\\
\eeq
It is convenient to see the potential as a function of $\Delta r\equiv r-\bar r_c$ and, even better, of $\Delta z\equiv\Delta r^3+i\Delta r^8$. Then
\beq
V_c(\Delta z;\mu)=V_c(e^{i\frac{2\pi}{3}k}\Delta z;\mu-i(2\pi/3) k T)\,.
\eeq
which implies
\beq
\Delta z(\mu)=e^{i\frac{2\pi}{3}k}\Delta z(\mu-i(2\pi/3)kT)\,.\label{eq:B13}
\eeq
Combining this with \eqref{eq:rCA}, we find that the phase of $\Delta z(i(\pi/3)kT)$ is $(\pi/3)k$ modulo $\pi$, corresponding to $r$ belonging to the symmetry axes of the Weyl chamber centered around $\bar r_c$. This is always true for even $k$ but can be broken for odd $k$, corresponding to the Roberge-Weiss transition already discussed in the main text.

Finally, the ${\cal K}$-transformation leads to
\beq
V_{\bar r_3,\bar r_8}(r_3,r_8;\mu)=V^*_{\bar r^*_3,-\bar r^*_8}(r^*_3,-r^*_8;\mu^*)\,,\label{eq:rK}
\eeq
while the ${\cal CK}$-transformation gives
\beq
V_{\bar r}(r;\mu)=V^*_{\bar r^*}(r^*;-\mu^*)\,.
\eeq
This second identity is useful in the case of an imaginary chemical potential. Indeed, in that case, and assuming a real-valued background (as is the case for $\bar r_c$), we find
\beq
V_{\bar r}(r;\mu)=V^*_{\bar r}(r^*;\mu)\,,
\eeq
from which it follows that
\beq
r^*(\mu)=r(\mu)\,.
\eeq
This suggests considering the potential for $\smash{r^*=r}$, that is for real-valued $r_3$ and $r_8$. In this case, the potential is real and the positivity of the fermion determinant allows one to argue that the stable extremum is the global minimum. Similarly, Eq.~\eqref{eq:rK} is useful in the case of a real chemical potential. Indeed, for any background with a vanishing $\bar r^8$ (at it is the case for $\bar r_c$), we find
\beq
V_{\bar r}(r_3,r_8;\mu)=V^*_{\bar r}(r^*_3,-r^*_8;\mu)\,,
\eeq
from which it follows that
\beq
r^*_3(\mu)=r_3(\mu) \quad {\rm and} \quad r_8^*(\mu)=-r_8(\mu)\,.
\eeq
This suggests considering the potential for real-valued $r_3$ and imaginary-valued $r_8$. In this case, the potential is also real. However, the fermionic determinant is not positive definite anymore, and the extrema, including the stable one, are usually saddle points.

All the constraints derived above on $r_3$ and $r_8$ map those derived in the main text for $\ell$ and $\bar\ell$ upon using Eqs.~\eqref{eq: l of r}-\eqref{eq: lb of r}.

\bibliography{HQreferences}

\begin{thebibliography}{67}%
\makeatletter
\providecommand \@ifxundefined [1]{%
 \@ifx{#1\undefined}
}%
\providecommand \@ifnum [1]{%
 \ifnum #1\expandafter \@firstoftwo
 \else \expandafter \@secondoftwo
 \fi
}%
\providecommand \@ifx [1]{%
 \ifx #1\expandafter \@firstoftwo
 \else \expandafter \@secondoftwo
 \fi
}%
\providecommand \natexlab [1]{#1}%
\providecommand \enquote  [1]{``#1''}%
\providecommand \bibnamefont  [1]{#1}%
\providecommand \bibfnamefont [1]{#1}%
\providecommand \citenamefont [1]{#1}%
\providecommand \href@noop [0]{\@secondoftwo}%
\providecommand \href [0]{\begingroup \@sanitize@url \@href}%
\providecommand \@href[1]{\@@startlink{#1}\@@href}%
\providecommand \@@href[1]{\endgroup#1\@@endlink}%
\providecommand \@sanitize@url [0]{\catcode `\\12\catcode `\$12\catcode `\&12\catcode `\#12\catcode `\^12\catcode `\_12\catcode `\%12\relax}%
\providecommand \@@startlink[1]{}%
\providecommand \@@endlink[0]{}%
\providecommand \url  [0]{\begingroup\@sanitize@url \@url }%
\providecommand \@url [1]{\endgroup\@href {#1}{\urlprefix }}%
\providecommand \urlprefix  [0]{URL }%
\providecommand \Eprint [0]{\href }%
\providecommand \doibase [0]{https://doi.org/}%
\providecommand \selectlanguage [0]{\@gobble}%
\providecommand \bibinfo  [0]{\@secondoftwo}%
\providecommand \bibfield  [0]{\@secondoftwo}%
\providecommand \translation [1]{[#1]}%
\providecommand \BibitemOpen [0]{}%
\providecommand \bibitemStop [0]{}%
\providecommand \bibitemNoStop [0]{.\EOS\space}%
\providecommand \EOS [0]{\spacefactor3000\relax}%
\providecommand \BibitemShut  [1]{\csname bibitem#1\endcsname}%
\let\auto@bib@innerbib\@empty
%</preamble>
\bibitem [{\citenamefont {Kolb}(2018)}]{Kolb2018TheUniverse}%
  \BibitemOpen
  \bibfield  {author} {\bibinfo {author} {\bibfnamefont {E.}~\bibnamefont {Kolb}},\ }\href {https://doi.org/10.1201/9780429492860} {\emph {\bibinfo {title} {{The Early Universe}}}},\ edited by\ \bibinfo {editor} {\bibfnamefont {E.~W.}\ \bibnamefont {Kolb}}\ and\ \bibinfo {editor} {\bibfnamefont {M.~S.}\ \bibnamefont {Turner}},\ Vol.~\bibinfo {volume} {69}\ (\bibinfo  {publisher} {CRC Press},\ \bibinfo {year} {2018})\BibitemShut {NoStop}%
\bibitem [{\citenamefont {Weissenborn}\ \emph {et~al.}(2011)\citenamefont {Weissenborn}, \citenamefont {Sagert}, \citenamefont {Pagliara}, \citenamefont {Hempel},\ and\ \citenamefont {Schaffner-Bielich}}]{Weissenborn2011QUARKSTARS}%
  \BibitemOpen
  \bibfield  {author} {\bibinfo {author} {\bibfnamefont {S.}~\bibnamefont {Weissenborn}}, \bibinfo {author} {\bibfnamefont {I.}~\bibnamefont {Sagert}}, \bibinfo {author} {\bibfnamefont {G.}~\bibnamefont {Pagliara}}, \bibinfo {author} {\bibfnamefont {M.}~\bibnamefont {Hempel}},\ and\ \bibinfo {author} {\bibfnamefont {J.}~\bibnamefont {Schaffner-Bielich}},\ }\href {https://doi.org/10.1088/2041-8205/740/1/L14} {\bibfield  {journal} {\bibinfo  {journal} {The Astrophysical Journal Letters}\ }\textbf {\bibinfo {volume} {740}},\ \bibinfo {pages} {L14} (\bibinfo {year} {2011})}\BibitemShut {NoStop}%
\bibitem [{\citenamefont {Kurkela}\ \emph {et~al.}(2014)\citenamefont {Kurkela}, \citenamefont {Fraga}, \citenamefont {Schaffner-Bielich},\ and\ \citenamefont {Vuorinen}}]{Kurkela2014CONSTRAININGCHROMODYNAMICS}%
  \BibitemOpen
  \bibfield  {author} {\bibinfo {author} {\bibfnamefont {A.}~\bibnamefont {Kurkela}}, \bibinfo {author} {\bibfnamefont {E.~S.}\ \bibnamefont {Fraga}}, \bibinfo {author} {\bibfnamefont {J.}~\bibnamefont {Schaffner-Bielich}},\ and\ \bibinfo {author} {\bibfnamefont {A.}~\bibnamefont {Vuorinen}},\ }\href {https://doi.org/10.1088/0004-637X/789/2/127} {\bibfield  {journal} {\bibinfo  {journal} {The Astrophysical Journal}\ }\textbf {\bibinfo {volume} {789}},\ \bibinfo {pages} {127} (\bibinfo {year} {2014})}\BibitemShut {NoStop}%
\bibitem [{\citenamefont {Borsanyi}\ \emph {et~al.}(2016)\citenamefont {Borsanyi}, \citenamefont {Fodor}, \citenamefont {Guenther}, \citenamefont {Kampert}, \citenamefont {Katz}, \citenamefont {Kawanai}, \citenamefont {Kovacs}, \citenamefont {Mages}, \citenamefont {Pasztor}, \citenamefont {Pittler}, \citenamefont {Redondo}, \citenamefont {Ringwald},\ and\ \citenamefont {Szabo}}]{Borsanyi2016CalculationChromodynamics}%
  \BibitemOpen
  \bibfield  {author} {\bibinfo {author} {\bibfnamefont {S.}~\bibnamefont {Borsanyi}}, \bibinfo {author} {\bibfnamefont {Z.}~\bibnamefont {Fodor}}, \bibinfo {author} {\bibfnamefont {J.}~\bibnamefont {Guenther}}, \bibinfo {author} {\bibfnamefont {K.-H.}\ \bibnamefont {Kampert}}, \bibinfo {author} {\bibfnamefont {S.~D.}\ \bibnamefont {Katz}}, \bibinfo {author} {\bibfnamefont {T.}~\bibnamefont {Kawanai}}, \bibinfo {author} {\bibfnamefont {T.~G.}\ \bibnamefont {Kovacs}}, \bibinfo {author} {\bibfnamefont {S.~W.}\ \bibnamefont {Mages}}, \bibinfo {author} {\bibfnamefont {A.}~\bibnamefont {Pasztor}}, \bibinfo {author} {\bibfnamefont {F.}~\bibnamefont {Pittler}}, \bibinfo {author} {\bibfnamefont {J.}~\bibnamefont {Redondo}}, \bibinfo {author} {\bibfnamefont {A.}~\bibnamefont {Ringwald}},\ and\ \bibinfo {author} {\bibfnamefont {K.~K.}\ \bibnamefont {Szabo}},\ }\href {https://doi.org/10.1038/nature20115} {\bibfield  {journal} {\bibinfo  {journal} {Nature}\ }\textbf {\bibinfo {volume} {539}},\ \bibinfo {pages} {69} (\bibinfo {year} {2016})}\BibitemShut {NoStop}%
\bibitem [{\citenamefont {Mohanty}(2011)}]{Mohanty2011STARRHIC}%
  \BibitemOpen
  \bibfield  {author} {\bibinfo {author} {\bibfnamefont {B.}~\bibnamefont {Mohanty}},\ }\href {https://doi.org/10.1088/0954-3899/38/12/124023} {\bibfield  {journal} {\bibinfo  {journal} {J. Phys. G:Nucl. Part. Phys.}\ }\textbf {\bibinfo {volume} {38}},\ \bibinfo {pages} {124023} (\bibinfo {year} {2011})}\BibitemShut {NoStop}%
\bibitem [{\citenamefont {{ALICE Collaboration}}(2024)}]{ALICE2024TheQCD}%
  \BibitemOpen
  \bibfield  {author} {\bibinfo {author} {\bibnamefont {{ALICE Collaboration}}},\ }\href {https://doi.org/10.1140/epjc/s10052-024-12935-y} {\bibfield  {journal} {\bibinfo  {journal} {Eur. Phys. J. C}\ }\textbf {\bibinfo {volume} {84}},\ \bibinfo {pages} {813} (\bibinfo {year} {2024})}\BibitemShut {NoStop}%
\bibitem [{\citenamefont {Mogliacci}\ \emph {et~al.}(2013)\citenamefont {Mogliacci}, \citenamefont {Andersen}, \citenamefont {Strickland}, \citenamefont {Su},\ and\ \citenamefont {Vuorinen}}]{Mogliacci:2013mca}%
  \BibitemOpen
  \bibfield  {author} {\bibinfo {author} {\bibfnamefont {S.}~\bibnamefont {Mogliacci}}, \bibinfo {author} {\bibfnamefont {J.~O.}\ \bibnamefont {Andersen}}, \bibinfo {author} {\bibfnamefont {M.}~\bibnamefont {Strickland}}, \bibinfo {author} {\bibfnamefont {N.}~\bibnamefont {Su}},\ and\ \bibinfo {author} {\bibfnamefont {A.}~\bibnamefont {Vuorinen}},\ }\href {https://doi.org/10.1007/JHEP12(2013)055} {\bibfield  {journal} {\bibinfo  {journal} {JHEP}\ }\textbf {\bibinfo {volume} {12}}\bibinfo  {number} { (2013)},\ \bibinfo {pages} {055}}\BibitemShut {NoStop}%
\bibitem [{\citenamefont {Bors{\'{a}}nyi}\ \emph {et~al.}(2014)\citenamefont {Bors{\'{a}}nyi}, \citenamefont {Fodor}, \citenamefont {Hoelbling}, \citenamefont {Katz}, \citenamefont {Krieg},\ and\ \citenamefont {Szab{\'{o}}}}]{Borsanyi2014FullFlavors}%
  \BibitemOpen
\bibfield  {number} {  }\bibfield  {author} {\bibinfo {author} {\bibfnamefont {S.}~\bibnamefont {Bors{\'{a}}nyi}}, \bibinfo {author} {\bibfnamefont {Z.}~\bibnamefont {Fodor}}, \bibinfo {author} {\bibfnamefont {C.}~\bibnamefont {Hoelbling}}, \bibinfo {author} {\bibfnamefont {S.~D.}\ \bibnamefont {Katz}}, \bibinfo {author} {\bibfnamefont {S.}~\bibnamefont {Krieg}},\ and\ \bibinfo {author} {\bibfnamefont {K.~K.}\ \bibnamefont {Szab{\'{o}}}},\ }\href {https://doi.org/10.1016/j.physletb.2014.01.007} {\bibfield  {journal} {\bibinfo  {journal} {Phys. Lett. B}\ }\textbf {\bibinfo {volume} {730}},\ \bibinfo {pages} {99} (\bibinfo {year} {2014})}\BibitemShut {NoStop}%
\bibitem [{\citenamefont {Bazavov}\ \emph {et~al.}(2014)\citenamefont {Bazavov}, \citenamefont {Bhattacharya}, \citenamefont {DeTar}, \citenamefont {Ding}, \citenamefont {Gottlieb}, \citenamefont {Gupta}, \citenamefont {Hegde}, \citenamefont {Heller}, \citenamefont {Karsch}, \citenamefont {Laermann}, \citenamefont {Levkova}, \citenamefont {Mukherjee}, \citenamefont {Petreczky}, \citenamefont {Schmidt}, \citenamefont {Schroeder}, \citenamefont {Soltz}, \citenamefont {Soeldner}, \citenamefont {Sugar}, \citenamefont {Wagner},\ and\ \citenamefont {Vranas}}]{Bazavov2014EquationQCD}%
  \BibitemOpen
  \bibfield  {author} {\bibinfo {author} {\bibfnamefont {A.}~\bibnamefont {Bazavov}}, \bibinfo {author} {\bibfnamefont {T.}~\bibnamefont {Bhattacharya}}, \bibinfo {author} {\bibfnamefont {C.}~\bibnamefont {DeTar}}, \bibinfo {author} {\bibfnamefont {H.-T.}\ \bibnamefont {Ding}}, \bibinfo {author} {\bibfnamefont {S.}~\bibnamefont {Gottlieb}}, \bibinfo {author} {\bibfnamefont {R.}~\bibnamefont {Gupta}}, \bibinfo {author} {\bibfnamefont {P.}~\bibnamefont {Hegde}}, \bibinfo {author} {\bibfnamefont {U.}~\bibnamefont {Heller}}, \bibinfo {author} {\bibfnamefont {F.}~\bibnamefont {Karsch}}, \bibinfo {author} {\bibfnamefont {E.}~\bibnamefont {Laermann}}, \bibinfo {author} {\bibfnamefont {L.}~\bibnamefont {Levkova}}, \bibinfo {author} {\bibfnamefont {S.}~\bibnamefont {Mukherjee}}, \bibinfo {author} {\bibfnamefont {P.}~\bibnamefont {Petreczky}}, \bibinfo {author} {\bibfnamefont {C.}~\bibnamefont {Schmidt}}, \bibinfo {author} {\bibfnamefont {C.}~\bibnamefont {Schroeder}}, \bibinfo {author} {\bibfnamefont {R.}~\bibnamefont {Soltz}}, \bibinfo {author} {\bibfnamefont {W.}~\bibnamefont {Soeldner}}, \bibinfo {author} {\bibfnamefont {R.}~\bibnamefont {Sugar}}, \bibinfo {author} {\bibfnamefont {M.}~\bibnamefont {Wagner}},\ and\ \bibinfo {author} {\bibfnamefont {P.}~\bibnamefont {Vranas}},\ }\href {https://doi.org/10.1103/PhysRevD.90.094503} {\bibfield  {journal} {\bibinfo  {journal} {Phys. Rev. D}\ }\textbf {\bibinfo {volume} {90}},\ \bibinfo {pages} {094503} (\bibinfo {year} {2014})}\BibitemShut {NoStop}%
\bibitem [{\citenamefont {Muroya}\ \emph {et~al.}(2003)\citenamefont {Muroya}, \citenamefont {Nakamura}, \citenamefont {Nonaka},\ and\ \citenamefont {Takaishi}}]{Muroya2003LatticeReview}%
  \BibitemOpen
  \bibfield  {author} {\bibinfo {author} {\bibfnamefont {S.}~\bibnamefont {Muroya}}, \bibinfo {author} {\bibfnamefont {A.}~\bibnamefont {Nakamura}}, \bibinfo {author} {\bibfnamefont {C.}~\bibnamefont {Nonaka}},\ and\ \bibinfo {author} {\bibfnamefont {T.}~\bibnamefont {Takaishi}},\ }\href {https://doi.org/10.1143/PTP.110.615} {\bibfield  {journal} {\bibinfo  {journal} {Prog. Theor. Phys.}\ }\textbf {\bibinfo {volume} {110}},\ \bibinfo {pages} {615} (\bibinfo {year} {2003})}\BibitemShut {NoStop}%
\bibitem [{\citenamefont {de~Forcrand}(2010)}]{deForcrand2010SimulatingDensity}%
  \BibitemOpen
  \bibfield  {author} {\bibinfo {author} {\bibfnamefont {P.}~\bibnamefont {de~Forcrand}},\ }in\ \href {https://doi.org/10.22323/1.091.0010} {\emph {\bibinfo {booktitle} {Proceedings of The XXVII International Symposium on Lattice Field Theory — PoS(LAT2009)}}}\ (\bibinfo  {publisher} {Sissa Medialab},\ \bibinfo {address} {Trieste, Italy},\ \bibinfo {year} {2010})\ p.\ \bibinfo {pages} {010}\BibitemShut {NoStop}%
\bibitem [{\citenamefont {Philipsen}(2012)}]{Philipsen2012LatticePotential}%
  \BibitemOpen
  \bibfield  {author} {\bibinfo {author} {\bibfnamefont {O.}~\bibnamefont {Philipsen}},\ }in\ \href {https://doi.org/10.22323/1.077.0011} {\emph {\bibinfo {booktitle} {Proceedings of VIIIth Conference Quark Confinement and the Hadron Spectrum — PoS(ConfinementVIII)}}}\ (\bibinfo  {publisher} {Sissa Medialab},\ \bibinfo {address} {Trieste, Italy},\ \bibinfo {year} {2012})\ p.\ \bibinfo {pages} {011}\BibitemShut {NoStop}%
\bibitem [{\citenamefont {Bors{\'{a}}nyi}\ \emph {et~al.}(2012)\citenamefont {Bors{\'{a}}nyi}, \citenamefont {Endr{\H{o}}di}, \citenamefont {Fodor}, \citenamefont {Katz}, \citenamefont {Krieg}, \citenamefont {Ratti},\ and\ \citenamefont {Szab{\'{o}}}}]{Borsanyi2012QCD2}%
  \BibitemOpen
  \bibfield  {author} {\bibinfo {author} {\bibfnamefont {S.}~\bibnamefont {Bors{\'{a}}nyi}}, \bibinfo {author} {\bibfnamefont {G.}~\bibnamefont {Endr{\H{o}}di}}, \bibinfo {author} {\bibfnamefont {Z.}~\bibnamefont {Fodor}}, \bibinfo {author} {\bibfnamefont {S.~D.}\ \bibnamefont {Katz}}, \bibinfo {author} {\bibfnamefont {S.}~\bibnamefont {Krieg}}, \bibinfo {author} {\bibfnamefont {C.}~\bibnamefont {Ratti}},\ and\ \bibinfo {author} {\bibfnamefont {K.~K.}\ \bibnamefont {Szab{\'{o}}}},\ }\href {https://doi.org/10.1007/JHEP08(2012)053} {\bibfield  {journal} {\bibinfo  {journal} {J. High Energy Phys.}\ }\textbf {\bibinfo {volume} {2012}}\bibinfo  {number} { (8)},\ \bibinfo {pages} {53}}\BibitemShut {NoStop}%
\bibitem [{\citenamefont {Nagata}(2022)}]{Nagata2022Finite-densityProblems}%
  \BibitemOpen
\bibfield  {number} {  }\bibfield  {author} {\bibinfo {author} {\bibfnamefont {K.}~\bibnamefont {Nagata}},\ }\href {https://doi.org/10.1016/j.ppnp.2022.103991} {\bibfield  {journal} {\bibinfo  {journal} {Prog. Part. Nucl. Phys.}\ }\textbf {\bibinfo {volume} {127}},\ \bibinfo {pages} {103991} (\bibinfo {year} {2022})}\BibitemShut {NoStop}%
\bibitem [{\citenamefont {Schaefer}\ and\ \citenamefont {Pirner}(1999)}]{Schaefer:1999em}%
  \BibitemOpen
  \bibfield  {author} {\bibinfo {author} {\bibfnamefont {B.-J.}\ \bibnamefont {Schaefer}}\ and\ \bibinfo {author} {\bibfnamefont {H.-J.}\ \bibnamefont {Pirner}},\ }\href {https://doi.org/10.1016/S0375-9474(99)00409-1} {\bibfield  {journal} {\bibinfo  {journal} {Nucl. Phys. A}\ }\textbf {\bibinfo {volume} {660}},\ \bibinfo {pages} {439} (\bibinfo {year} {1999})}\BibitemShut {NoStop}%
\bibitem [{\citenamefont {Adhikari}\ \emph {et~al.}(2018)\citenamefont {Adhikari}, \citenamefont {Andersen},\ and\ \citenamefont {Kneschke}}]{Adhikari2018PionModel}%
  \BibitemOpen
  \bibfield  {author} {\bibinfo {author} {\bibfnamefont {P.}~\bibnamefont {Adhikari}}, \bibinfo {author} {\bibfnamefont {J.~O.}\ \bibnamefont {Andersen}},\ and\ \bibinfo {author} {\bibfnamefont {P.}~\bibnamefont {Kneschke}},\ }\href {https://doi.org/10.1103/PhysRevD.98.074016} {\bibfield  {journal} {\bibinfo  {journal} {Phys. Rev. D}\ }\textbf {\bibinfo {volume} {98}},\ \bibinfo {pages} {074016} (\bibinfo {year} {2018})}\BibitemShut {NoStop}%
\bibitem [{\citenamefont {Kov{\'{a}}cs}\ \emph {et~al.}(2016)\citenamefont {Kov{\'{a}}cs}, \citenamefont {Sz{\'{e}}p},\ and\ \citenamefont {Wolf}}]{Kovacs2016ExistenceModel}%
  \BibitemOpen
  \bibfield  {author} {\bibinfo {author} {\bibfnamefont {P.}~\bibnamefont {Kov{\'{a}}cs}}, \bibinfo {author} {\bibfnamefont {Z.}~\bibnamefont {Sz{\'{e}}p}},\ and\ \bibinfo {author} {\bibfnamefont {G.}~\bibnamefont {Wolf}},\ }\href {https://doi.org/10.1103/PhysRevD.93.114014} {\bibfield  {journal} {\bibinfo  {journal} {Phys. Rev. D}\ }\textbf {\bibinfo {volume} {93}},\ \bibinfo {pages} {114014} (\bibinfo {year} {2016})}\BibitemShut {NoStop}%
\bibitem [{\citenamefont {Fukushima}(2004)}]{Fukushima:2003fw}%
  \BibitemOpen
  \bibfield  {author} {\bibinfo {author} {\bibfnamefont {K.}~\bibnamefont {Fukushima}},\ }\href {https://doi.org/10.1016/j.physletb.2004.04.027} {\bibfield  {journal} {\bibinfo  {journal} {Phys. Lett. B}\ }\textbf {\bibinfo {volume} {591}},\ \bibinfo {pages} {277} (\bibinfo {year} {2004})}\BibitemShut {NoStop}%
\bibitem [{\citenamefont {Ratti}\ \emph {et~al.}(2006)\citenamefont {Ratti}, \citenamefont {Thaler},\ and\ \citenamefont {Weise}}]{Ratti:2005jh}%
  \BibitemOpen
  \bibfield  {author} {\bibinfo {author} {\bibfnamefont {C.}~\bibnamefont {Ratti}}, \bibinfo {author} {\bibfnamefont {M.~A.}\ \bibnamefont {Thaler}},\ and\ \bibinfo {author} {\bibfnamefont {W.}~\bibnamefont {Weise}},\ }\href {https://doi.org/10.1103/PhysRevD.73.014019} {\bibfield  {journal} {\bibinfo  {journal} {Phys. Rev. D}\ }\textbf {\bibinfo {volume} {73}},\ \bibinfo {pages} {014019} (\bibinfo {year} {2006})}\BibitemShut {NoStop}%
\bibitem [{\citenamefont {Roessner}\ \emph {et~al.}(2007)\citenamefont {Roessner}, \citenamefont {Ratti},\ and\ \citenamefont {Weise}}]{Roessner:2006xn}%
  \BibitemOpen
  \bibfield  {author} {\bibinfo {author} {\bibfnamefont {S.}~\bibnamefont {Roessner}}, \bibinfo {author} {\bibfnamefont {C.}~\bibnamefont {Ratti}},\ and\ \bibinfo {author} {\bibfnamefont {W.}~\bibnamefont {Weise}},\ }\href {https://doi.org/10.1103/PhysRevD.75.034007} {\bibfield  {journal} {\bibinfo  {journal} {Phys. Rev. D}\ }\textbf {\bibinfo {volume} {75}},\ \bibinfo {pages} {034007} (\bibinfo {year} {2007})}\BibitemShut {NoStop}%
\bibitem [{\citenamefont {Fujihara}\ \emph {et~al.}(2009)\citenamefont {Fujihara}, \citenamefont {Kimura}, \citenamefont {Inagaki},\ and\ \citenamefont {Kvinikhidze}}]{Fujihara:2008ae}%
  \BibitemOpen
  \bibfield  {author} {\bibinfo {author} {\bibfnamefont {T.}~\bibnamefont {Fujihara}}, \bibinfo {author} {\bibfnamefont {D.}~\bibnamefont {Kimura}}, \bibinfo {author} {\bibfnamefont {T.}~\bibnamefont {Inagaki}},\ and\ \bibinfo {author} {\bibfnamefont {A.}~\bibnamefont {Kvinikhidze}},\ }\href {https://doi.org/10.1103/PhysRevD.79.096008} {\bibfield  {journal} {\bibinfo  {journal} {Phys. Rev. D}\ }\textbf {\bibinfo {volume} {79}},\ \bibinfo {pages} {096008} (\bibinfo {year} {2009})}\BibitemShut {NoStop}%
\bibitem [{\citenamefont {Fischer}(2019)}]{Fischer2019QCDEquations}%
  \BibitemOpen
  \bibfield  {author} {\bibinfo {author} {\bibfnamefont {C.~S.}\ \bibnamefont {Fischer}},\ }\href {https://doi.org/10.1016/j.ppnp.2019.01.002} {\bibfield  {journal} {\bibinfo  {journal} {Prog. Part. Nucl. Phys.}\ }\textbf {\bibinfo {volume} {105}},\ \bibinfo {pages} {1} (\bibinfo {year} {2019})}\BibitemShut {NoStop}%
\bibitem [{\citenamefont {Fu}\ \emph {et~al.}(2020)\citenamefont {Fu}, \citenamefont {Pawlowski},\ and\ \citenamefont {Rennecke}}]{Fu2020QCDDensity}%
  \BibitemOpen
  \bibfield  {author} {\bibinfo {author} {\bibfnamefont {W.-j.}\ \bibnamefont {Fu}}, \bibinfo {author} {\bibfnamefont {J.~M.}\ \bibnamefont {Pawlowski}},\ and\ \bibinfo {author} {\bibfnamefont {F.}~\bibnamefont {Rennecke}},\ }\href {https://doi.org/10.1103/PhysRevD.101.054032} {\bibfield  {journal} {\bibinfo  {journal} {Phys. Rev. D}\ }\textbf {\bibinfo {volume} {101}},\ \bibinfo {pages} {054032} (\bibinfo {year} {2020})}\BibitemShut {NoStop}%
\bibitem [{\citenamefont {Gribov}(1978)}]{Gribov1978QuantizationTheories}%
  \BibitemOpen
  \bibfield  {author} {\bibinfo {author} {\bibfnamefont {V.~N.}\ \bibnamefont {Gribov}},\ }\href {https://doi.org/10.1016/0550-3213(78)90175-X} {\bibfield  {journal} {\bibinfo  {journal} {Nucl. Phys. B}\ }\textbf {\bibinfo {volume} {139}},\ \bibinfo {pages} {1} (\bibinfo {year} {1978})}\BibitemShut {NoStop}%
\bibitem [{\citenamefont {Singer}(1978)}]{Singer1978SomeAmbiguity}%
  \BibitemOpen
  \bibfield  {author} {\bibinfo {author} {\bibfnamefont {I.~M.}\ \bibnamefont {Singer}},\ }\href {https://doi.org/10.1007/BF01609471} {\bibfield  {journal} {\bibinfo  {journal} {Commun. Math. Phys.}\ }\textbf {\bibinfo {volume} {60}},\ \bibinfo {pages} {7} (\bibinfo {year} {1978})}\BibitemShut {NoStop}%
\bibitem [{\citenamefont {Eichmann}\ \emph {et~al.}(2016)\citenamefont {Eichmann}, \citenamefont {Sanchis-Alepuz}, \citenamefont {Williams}, \citenamefont {Alkofer},\ and\ \citenamefont {Fischer}}]{Eichmann2016BaryonsStates}%
  \BibitemOpen
  \bibfield  {author} {\bibinfo {author} {\bibfnamefont {G.}~\bibnamefont {Eichmann}}, \bibinfo {author} {\bibfnamefont {H.}~\bibnamefont {Sanchis-Alepuz}}, \bibinfo {author} {\bibfnamefont {R.}~\bibnamefont {Williams}}, \bibinfo {author} {\bibfnamefont {R.}~\bibnamefont {Alkofer}},\ and\ \bibinfo {author} {\bibfnamefont {C.~S.}\ \bibnamefont {Fischer}},\ }\href {https://doi.org/10.1016/j.ppnp.2016.07.001} {\bibfield  {journal} {\bibinfo  {journal} {Progress in Particle and Nuclear Physics}\ }\textbf {\bibinfo {volume} {91}},\ \bibinfo {pages} {1} (\bibinfo {year} {2016})}\BibitemShut {NoStop}%
\bibitem [{\citenamefont {Huber}(2020)}]{Huber2020NonperturbativeTheories}%
  \BibitemOpen
  \bibfield  {author} {\bibinfo {author} {\bibfnamefont {M.~Q.}\ \bibnamefont {Huber}},\ }\href {https://doi.org/10.1016/j.physrep.2020.04.004} {\bibfield  {journal} {\bibinfo  {journal} {Physics Reports}\ }\textbf {\bibinfo {volume} {879}},\ \bibinfo {pages} {1} (\bibinfo {year} {2020})}\BibitemShut {NoStop}%
\bibitem [{\citenamefont {Curci}\ and\ \citenamefont {Ferrari}(1976)}]{Curci1976OnFields}%
  \BibitemOpen
  \bibfield  {author} {\bibinfo {author} {\bibfnamefont {G.}~\bibnamefont {Curci}}\ and\ \bibinfo {author} {\bibfnamefont {R.}~\bibnamefont {Ferrari}},\ }\href {https://doi.org/10.1007/BF02729999} {\bibfield  {journal} {\bibinfo  {journal} {Il Nuovo Cimento A}\ }\textbf {\bibinfo {volume} {32}},\ \bibinfo {pages} {151} (\bibinfo {year} {1976})}\BibitemShut {NoStop}%
\bibitem [{\citenamefont {Bonnet}\ \emph {et~al.}(2000)\citenamefont {Bonnet}, \citenamefont {Bowman}, \citenamefont {Leinweber},\ and\ \citenamefont {Williams}}]{Bonnet2000InfraredLattice}%
  \BibitemOpen
  \bibfield  {author} {\bibinfo {author} {\bibfnamefont {F.~D.~R.}\ \bibnamefont {Bonnet}}, \bibinfo {author} {\bibfnamefont {P.~O.}\ \bibnamefont {Bowman}}, \bibinfo {author} {\bibfnamefont {D.~B.}\ \bibnamefont {Leinweber}},\ and\ \bibinfo {author} {\bibfnamefont {A.~G.}\ \bibnamefont {Williams}},\ }\href {https://doi.org/10.1103/PhysRevD.62.051501} {\bibfield  {journal} {\bibinfo  {journal} {Phys. Rev. D}\ }\textbf {\bibinfo {volume} {62}},\ \bibinfo {pages} {051501} (\bibinfo {year} {2000})}\BibitemShut {NoStop}%
\bibitem [{\citenamefont {Bonnet}\ \emph {et~al.}(2001)\citenamefont {Bonnet}, \citenamefont {Bowman}, \citenamefont {Leinweber}, \citenamefont {Williams},\ and\ \citenamefont {Zanotti}}]{Bonnet2001InfinitePropagator}%
  \BibitemOpen
  \bibfield  {author} {\bibinfo {author} {\bibfnamefont {F.~D.~R.}\ \bibnamefont {Bonnet}}, \bibinfo {author} {\bibfnamefont {P.~O.}\ \bibnamefont {Bowman}}, \bibinfo {author} {\bibfnamefont {D.~B.}\ \bibnamefont {Leinweber}}, \bibinfo {author} {\bibfnamefont {A.~G.}\ \bibnamefont {Williams}},\ and\ \bibinfo {author} {\bibfnamefont {J.~M.}\ \bibnamefont {Zanotti}},\ }\href {https://doi.org/10.1103/PhysRevD.64.034501} {\bibfield  {journal} {\bibinfo  {journal} {Phys. Rev. D}\ }\textbf {\bibinfo {volume} {64}},\ \bibinfo {pages} {034501} (\bibinfo {year} {2001})}\BibitemShut {NoStop}%
\bibitem [{\citenamefont {Cucchieri}\ and\ \citenamefont {Mendes}(2008)}]{Cucchieri2008ConstraintsTheories}%
  \BibitemOpen
  \bibfield  {author} {\bibinfo {author} {\bibfnamefont {A.}~\bibnamefont {Cucchieri}}\ and\ \bibinfo {author} {\bibfnamefont {T.}~\bibnamefont {Mendes}},\ }\href {https://doi.org/10.1103/PhysRevLett.100.241601} {\bibfield  {journal} {\bibinfo  {journal} {Phys. Rev. Lett.}\ }\textbf {\bibinfo {volume} {100}},\ \bibinfo {pages} {241601} (\bibinfo {year} {2008})}\BibitemShut {NoStop}%
\bibitem [{\citenamefont {Bogolubsky}\ \emph {et~al.}(2009)\citenamefont {Bogolubsky}, \citenamefont {Ilgenfritz}, \citenamefont {M{\"{u}}ller-Preussker},\ and\ \citenamefont {Sternbeck}}]{Bogolubsky2009LatticeInfrared}%
  \BibitemOpen
  \bibfield  {author} {\bibinfo {author} {\bibfnamefont {I.~L.}\ \bibnamefont {Bogolubsky}}, \bibinfo {author} {\bibfnamefont {E.-M.}\ \bibnamefont {Ilgenfritz}}, \bibinfo {author} {\bibfnamefont {M.}~\bibnamefont {M{\"{u}}ller-Preussker}},\ and\ \bibinfo {author} {\bibfnamefont {A.}~\bibnamefont {Sternbeck}},\ }\href {https://doi.org/10.1016/j.physletb.2009.04.076} {\bibfield  {journal} {\bibinfo  {journal} {Phys. Lett. B}\ }\textbf {\bibinfo {volume} {676}},\ \bibinfo {pages} {69} (\bibinfo {year} {2009})}\BibitemShut {NoStop}%
\bibitem [{\citenamefont {Oliveira}\ and\ \citenamefont {Silva}(2012)}]{Oliveira2012LatticeDependence}%
  \BibitemOpen
  \bibfield  {author} {\bibinfo {author} {\bibfnamefont {O.}~\bibnamefont {Oliveira}}\ and\ \bibinfo {author} {\bibfnamefont {P.~J.}\ \bibnamefont {Silva}},\ }\href {https://doi.org/10.1103/PhysRevD.86.114513} {\bibfield  {journal} {\bibinfo  {journal} {Phys. Rev. D}\ }\textbf {\bibinfo {volume} {86}},\ \bibinfo {pages} {114513} (\bibinfo {year} {2012})}\BibitemShut {NoStop}%
\bibitem [{\citenamefont {Maas}(2013)}]{Maas2013GaugeTemperature}%
  \BibitemOpen
  \bibfield  {author} {\bibinfo {author} {\bibfnamefont {A.}~\bibnamefont {Maas}},\ }\href {https://doi.org/10.1016/j.physrep.2012.11.002} {\bibfield  {journal} {\bibinfo  {journal} {Phys. Rep.}\ }\textbf {\bibinfo {volume} {524}},\ \bibinfo {pages} {203} (\bibinfo {year} {2013})}\BibitemShut {NoStop}%
\bibitem [{\citenamefont {Silva}\ \emph {et~al.}(2014)\citenamefont {Silva}, \citenamefont {Oliveira}, \citenamefont {Bicudo},\ and\ \citenamefont {Cardoso}}]{Silva2014GluonQCD}%
  \BibitemOpen
  \bibfield  {author} {\bibinfo {author} {\bibfnamefont {P.}~\bibnamefont {Silva}}, \bibinfo {author} {\bibfnamefont {O.}~\bibnamefont {Oliveira}}, \bibinfo {author} {\bibfnamefont {P.}~\bibnamefont {Bicudo}},\ and\ \bibinfo {author} {\bibfnamefont {N.}~\bibnamefont {Cardoso}},\ }\href {https://doi.org/10.1103/PhysRevD.89.074503} {\bibfield  {journal} {\bibinfo  {journal} {Phys. Rev. D}\ }\textbf {\bibinfo {volume} {89}},\ \bibinfo {pages} {074503} (\bibinfo {year} {2014})}\BibitemShut {NoStop}%
\bibitem [{\citenamefont {Skullerud}\ \emph {et~al.}(2003)\citenamefont {Skullerud}, \citenamefont {Bowman}, \citenamefont {Kizilersu}, \citenamefont {Leinweber},\ and\ \citenamefont {Williams}}]{Skullerud:2003qu}%
  \BibitemOpen
  \bibfield  {author} {\bibinfo {author} {\bibfnamefont {J.~I.}\ \bibnamefont {Skullerud}}, \bibinfo {author} {\bibfnamefont {P.~O.}\ \bibnamefont {Bowman}}, \bibinfo {author} {\bibfnamefont {A.}~\bibnamefont {Kizilersu}}, \bibinfo {author} {\bibfnamefont {D.~B.}\ \bibnamefont {Leinweber}},\ and\ \bibinfo {author} {\bibfnamefont {A.~G.}\ \bibnamefont {Williams}},\ }\href {https://doi.org/10.1088/1126-6708/2003/04/047} {\bibfield  {journal} {\bibinfo  {journal} {JHEP}\ }\textbf {\bibinfo {volume} {04}}\bibinfo  {number} { (2003)},\ \bibinfo {pages} {047}}\BibitemShut {NoStop}%
\bibitem [{\citenamefont {Tissier}\ and\ \citenamefont {Wschebor}(2010)}]{Tissier2010InfraredTheory}%
  \BibitemOpen
\bibfield  {number} {  }\bibfield  {author} {\bibinfo {author} {\bibfnamefont {M.}~\bibnamefont {Tissier}}\ and\ \bibinfo {author} {\bibfnamefont {N.}~\bibnamefont {Wschebor}},\ }\bibfield  {journal} {\bibinfo  {journal} {Phys. Rev. D}\ }\textbf {\bibinfo {volume} {82}},\ \href {https://doi.org/10.1103/PhysRevD.82.101701} {10.1103/PhysRevD.82.101701} (\bibinfo {year} {2010})\BibitemShut {NoStop}%
\bibitem [{\citenamefont {Tissier}\ and\ \citenamefont {Wschebor}(2011)}]{Tissier2011InfraredCorrelators}%
  \BibitemOpen
  \bibfield  {author} {\bibinfo {author} {\bibfnamefont {M.}~\bibnamefont {Tissier}}\ and\ \bibinfo {author} {\bibfnamefont {N.}~\bibnamefont {Wschebor}},\ }\bibfield  {journal} {\bibinfo  {journal} {Phys. Rev. D}\ }\textbf {\bibinfo {volume} {84}},\ \href {https://doi.org/10.1103/PhysRevD.84.045018} {10.1103/PhysRevD.84.045018} (\bibinfo {year} {2011})\BibitemShut {NoStop}%
\bibitem [{\citenamefont {Pel{\'{a}}ez}\ \emph {et~al.}(2013)\citenamefont {Pel{\'{a}}ez}, \citenamefont {Tissier},\ and\ \citenamefont {Wschebor}}]{Pelaez2013Three-pointTheory}%
  \BibitemOpen
  \bibfield  {author} {\bibinfo {author} {\bibfnamefont {M.}~\bibnamefont {Pel{\'{a}}ez}}, \bibinfo {author} {\bibfnamefont {M.}~\bibnamefont {Tissier}},\ and\ \bibinfo {author} {\bibfnamefont {N.}~\bibnamefont {Wschebor}},\ }\href {https://doi.org/10.1103/PhysRevD.88.125003} {\bibfield  {journal} {\bibinfo  {journal} {Phys. Rev. D}\ }\textbf {\bibinfo {volume} {88}},\ \bibinfo {pages} {125003} (\bibinfo {year} {2013})}\BibitemShut {NoStop}%
\bibitem [{\citenamefont {Pel{\'{a}}ez}\ \emph {et~al.}(2014)\citenamefont {Pel{\'{a}}ez}, \citenamefont {Tissier},\ and\ \citenamefont {Wschebor}}]{Pelaez2014Two-pointGauge}%
  \BibitemOpen
  \bibfield  {author} {\bibinfo {author} {\bibfnamefont {M.}~\bibnamefont {Pel{\'{a}}ez}}, \bibinfo {author} {\bibfnamefont {M.}~\bibnamefont {Tissier}},\ and\ \bibinfo {author} {\bibfnamefont {N.}~\bibnamefont {Wschebor}},\ }\href {https://doi.org/10.1103/PhysRevD.90.065031} {\bibfield  {journal} {\bibinfo  {journal} {Phys. Rev. D}\ }\textbf {\bibinfo {volume} {90}},\ \bibinfo {pages} {065031} (\bibinfo {year} {2014})}\BibitemShut {NoStop}%
\bibitem [{\citenamefont {Pel{\'{a}}ez}\ \emph {et~al.}(2015)\citenamefont {Pel{\'{a}}ez}, \citenamefont {Tissier},\ and\ \citenamefont {Wschebor}}]{Pelaez2015Quark-gluonModel}%
  \BibitemOpen
  \bibfield  {author} {\bibinfo {author} {\bibfnamefont {M.}~\bibnamefont {Pel{\'{a}}ez}}, \bibinfo {author} {\bibfnamefont {M.}~\bibnamefont {Tissier}},\ and\ \bibinfo {author} {\bibfnamefont {N.}~\bibnamefont {Wschebor}},\ }\href {https://doi.org/10.1103/PhysRevD.92.045012} {\bibfield  {journal} {\bibinfo  {journal} {Phys. Rev. D}\ }\textbf {\bibinfo {volume} {92}},\ \bibinfo {pages} {045012} (\bibinfo {year} {2015})}\BibitemShut {NoStop}%
\bibitem [{\citenamefont {Reinosa}\ \emph {et~al.}(2014)\citenamefont {Reinosa}, \citenamefont {Serreau}, \citenamefont {Tissier},\ and\ \citenamefont {Wschebor}}]{Reinosa2014Yang-MillsPerspective}%
  \BibitemOpen
  \bibfield  {author} {\bibinfo {author} {\bibfnamefont {U.}~\bibnamefont {Reinosa}}, \bibinfo {author} {\bibfnamefont {J.}~\bibnamefont {Serreau}}, \bibinfo {author} {\bibfnamefont {M.}~\bibnamefont {Tissier}},\ and\ \bibinfo {author} {\bibfnamefont {N.}~\bibnamefont {Wschebor}},\ }\href {https://doi.org/10.1103/PhysRevD.89.105016} {\bibfield  {journal} {\bibinfo  {journal} {Phys. Rev. D}\ }\textbf {\bibinfo {volume} {89}},\ \bibinfo {pages} {105016} (\bibinfo {year} {2014})}\BibitemShut {NoStop}%
\bibitem [{\citenamefont {Pel\'aez}\ \emph {et~al.}(2017)\citenamefont {Pel\'aez}, \citenamefont {Reinosa}, \citenamefont {Serreau}, \citenamefont {Tissier},\ and\ \citenamefont {Wschebor}}]{Pelaez:2017bhh}%
  \BibitemOpen
  \bibfield  {author} {\bibinfo {author} {\bibfnamefont {M.}~\bibnamefont {Pel\'aez}}, \bibinfo {author} {\bibfnamefont {U.}~\bibnamefont {Reinosa}}, \bibinfo {author} {\bibfnamefont {J.}~\bibnamefont {Serreau}}, \bibinfo {author} {\bibfnamefont {M.}~\bibnamefont {Tissier}},\ and\ \bibinfo {author} {\bibfnamefont {N.}~\bibnamefont {Wschebor}},\ }\href {https://doi.org/10.1103/PhysRevD.96.114011} {\bibfield  {journal} {\bibinfo  {journal} {Phys. Rev. D}\ }\textbf {\bibinfo {volume} {96}},\ \bibinfo {pages} {114011} (\bibinfo {year} {2017})}\BibitemShut {NoStop}%
\bibitem [{\citenamefont {Pel\'aez}\ \emph {et~al.}(2021)\citenamefont {Pel\'aez}, \citenamefont {Reinosa}, \citenamefont {Serreau}, \citenamefont {Tissier},\ and\ \citenamefont {Wschebor}}]{Pelaez:2020ups}%
  \BibitemOpen
  \bibfield  {author} {\bibinfo {author} {\bibfnamefont {M.}~\bibnamefont {Pel\'aez}}, \bibinfo {author} {\bibfnamefont {U.}~\bibnamefont {Reinosa}}, \bibinfo {author} {\bibfnamefont {J.}~\bibnamefont {Serreau}}, \bibinfo {author} {\bibfnamefont {M.}~\bibnamefont {Tissier}},\ and\ \bibinfo {author} {\bibfnamefont {N.}~\bibnamefont {Wschebor}},\ }\href {https://doi.org/10.1103/PhysRevD.103.094035} {\bibfield  {journal} {\bibinfo  {journal} {Phys. Rev. D}\ }\textbf {\bibinfo {volume} {103}},\ \bibinfo {pages} {094035} (\bibinfo {year} {2021})}\BibitemShut {NoStop}%
\bibitem [{\citenamefont {Pel{\'{a}}ez}\ \emph {et~al.}(2021)\citenamefont {Pel{\'{a}}ez}, \citenamefont {Reinosa}, \citenamefont {Serreau}, \citenamefont {Tissier},\ and\ \citenamefont {Wschebor}}]{Pelaez2021AParameters}%
  \BibitemOpen
  \bibfield  {author} {\bibinfo {author} {\bibfnamefont {M.}~\bibnamefont {Pel{\'{a}}ez}}, \bibinfo {author} {\bibfnamefont {U.}~\bibnamefont {Reinosa}}, \bibinfo {author} {\bibfnamefont {J.}~\bibnamefont {Serreau}}, \bibinfo {author} {\bibfnamefont {M.}~\bibnamefont {Tissier}},\ and\ \bibinfo {author} {\bibfnamefont {N.}~\bibnamefont {Wschebor}},\ }\href {https://doi.org/10.1088/1361-6633/ac36b8} {\bibfield  {journal} {\bibinfo  {journal} {Rep. Prog. Phys.}\ }\textbf {\bibinfo {volume} {84}},\ \bibinfo {pages} {124202} (\bibinfo {year} {2021})}\BibitemShut {NoStop}%
\bibitem [{\citenamefont {Maelger}\ \emph {et~al.}(2020)\citenamefont {Maelger}, \citenamefont {Reinosa},\ and\ \citenamefont {Serreau}}]{Maelger:2019cbk}%
  \BibitemOpen
  \bibfield  {author} {\bibinfo {author} {\bibfnamefont {J.}~\bibnamefont {Maelger}}, \bibinfo {author} {\bibfnamefont {U.}~\bibnamefont {Reinosa}},\ and\ \bibinfo {author} {\bibfnamefont {J.}~\bibnamefont {Serreau}},\ }\href {https://doi.org/10.1103/PhysRevD.101.014028} {\bibfield  {journal} {\bibinfo  {journal} {Phys. Rev. D}\ }\textbf {\bibinfo {volume} {101}},\ \bibinfo {pages} {014028} (\bibinfo {year} {2020})}\BibitemShut {NoStop}%
\bibitem [{\citenamefont {DeWitt}(1967)}]{DeWitt1967QuantumTheory}%
  \BibitemOpen
  \bibfield  {author} {\bibinfo {author} {\bibfnamefont {B.~S.}\ \bibnamefont {DeWitt}},\ }\href {https://doi.org/10.1103/PhysRev.162.1195} {\bibfield  {journal} {\bibinfo  {journal} {Phys. Rev.}\ }\textbf {\bibinfo {volume} {162}},\ \bibinfo {pages} {1195} (\bibinfo {year} {1967})}\BibitemShut {NoStop}%
\bibitem [{\citenamefont {Polyakov}(1978)}]{Polyakov1978ThermalLiberation}%
  \BibitemOpen
  \bibfield  {author} {\bibinfo {author} {\bibfnamefont {A.~M.}\ \bibnamefont {Polyakov}},\ }\href {https://doi.org/10.1016/0370-2693(78)90737-2} {\bibfield  {journal} {\bibinfo  {journal} {Phys. Lett. B}\ }\textbf {\bibinfo {volume} {72}},\ \bibinfo {pages} {477} (\bibinfo {year} {1978})}\BibitemShut {NoStop}%
\bibitem [{\citenamefont {Braun}\ \emph {et~al.}(2010)\citenamefont {Braun}, \citenamefont {Gies},\ and\ \citenamefont {Pawlowski}}]{Braun2010QuarkConfinement}%
  \BibitemOpen
  \bibfield  {author} {\bibinfo {author} {\bibfnamefont {J.}~\bibnamefont {Braun}}, \bibinfo {author} {\bibfnamefont {H.}~\bibnamefont {Gies}},\ and\ \bibinfo {author} {\bibfnamefont {J.~M.}\ \bibnamefont {Pawlowski}},\ }\href {https://doi.org/10.1016/J.PHYSLETB.2010.01.009} {\bibfield  {journal} {\bibinfo  {journal} {Phys. Lett. B}\ }\textbf {\bibinfo {volume} {684}},\ \bibinfo {pages} {262} (\bibinfo {year} {2010})}\BibitemShut {NoStop}%
\bibitem [{\citenamefont {Reinosa}\ \emph {et~al.}(2015{\natexlab{a}})\citenamefont {Reinosa}, \citenamefont {Serreau}, \citenamefont {Tissier},\ and\ \citenamefont {Wschebor}}]{Reinosa2015DeconfinementTheory}%
  \BibitemOpen
  \bibfield  {author} {\bibinfo {author} {\bibfnamefont {U.}~\bibnamefont {Reinosa}}, \bibinfo {author} {\bibfnamefont {J.}~\bibnamefont {Serreau}}, \bibinfo {author} {\bibfnamefont {M.}~\bibnamefont {Tissier}},\ and\ \bibinfo {author} {\bibfnamefont {N.}~\bibnamefont {Wschebor}},\ }\href {https://doi.org/10.1016/J.PHYSLETB.2015.01.006} {\bibfield  {journal} {\bibinfo  {journal} {Phys. Lett. B}\ }\textbf {\bibinfo {volume} {742}},\ \bibinfo {pages} {61} (\bibinfo {year} {2015}{\natexlab{a}})}\BibitemShut {NoStop}%
\bibitem [{\citenamefont {Reinosa}\ \emph {et~al.}(2015{\natexlab{b}})\citenamefont {Reinosa}, \citenamefont {Serreau},\ and\ \citenamefont {Tissier}}]{Reinosa2015PerturbativePotential}%
  \BibitemOpen
  \bibfield  {author} {\bibinfo {author} {\bibfnamefont {U.}~\bibnamefont {Reinosa}}, \bibinfo {author} {\bibfnamefont {J.}~\bibnamefont {Serreau}},\ and\ \bibinfo {author} {\bibfnamefont {M.}~\bibnamefont {Tissier}},\ }\href {https://doi.org/10.1103/PhysRevD.92.025021} {\bibfield  {journal} {\bibinfo  {journal} {Phys. Rev. D}\ }\textbf {\bibinfo {volume} {92}},\ \bibinfo {pages} {025021} (\bibinfo {year} {2015}{\natexlab{b}})}\BibitemShut {NoStop}%
\bibitem [{\citenamefont {Reinosa}\ \emph {et~al.}(2016)\citenamefont {Reinosa}, \citenamefont {Serreau}, \citenamefont {Tissier},\ and\ \citenamefont {Wschebor}}]{Reinosa2016Two-loopBeyond}%
  \BibitemOpen
  \bibfield  {author} {\bibinfo {author} {\bibfnamefont {U.}~\bibnamefont {Reinosa}}, \bibinfo {author} {\bibfnamefont {J.}~\bibnamefont {Serreau}}, \bibinfo {author} {\bibfnamefont {M.}~\bibnamefont {Tissier}},\ and\ \bibinfo {author} {\bibfnamefont {N.}~\bibnamefont {Wschebor}},\ }\href {https://doi.org/10.1103/PhysRevD.93.105002} {\bibfield  {journal} {\bibinfo  {journal} {Phys. Rev. D}\ }\textbf {\bibinfo {volume} {93}},\ \bibinfo {pages} {105002} (\bibinfo {year} {2016})}\BibitemShut {NoStop}%
\bibitem [{\citenamefont {Maelger}\ \emph {et~al.}(2018{\natexlab{a}})\citenamefont {Maelger}, \citenamefont {Reinosa},\ and\ \citenamefont {Serreau}}]{Maelger2018PerturbativeCorrections}%
  \BibitemOpen
  \bibfield  {author} {\bibinfo {author} {\bibfnamefont {J.}~\bibnamefont {Maelger}}, \bibinfo {author} {\bibfnamefont {U.}~\bibnamefont {Reinosa}},\ and\ \bibinfo {author} {\bibfnamefont {J.}~\bibnamefont {Serreau}},\ }\href {https://doi.org/10.1103/PhysRevD.97.074027} {\bibfield  {journal} {\bibinfo  {journal} {Phys. Rev. D}\ }\textbf {\bibinfo {volume} {97}},\ \bibinfo {pages} {074027} (\bibinfo {year} {2018}{\natexlab{a}})}\BibitemShut {NoStop}%
\bibitem [{\citenamefont {van Egmond}\ \emph {et~al.}(2022)\citenamefont {van Egmond}, \citenamefont {Reinosa}, \citenamefont {Serreau},\ and\ \citenamefont {Tissier}}]{vanEgmond2022ATransition}%
  \BibitemOpen
  \bibfield  {author} {\bibinfo {author} {\bibfnamefont {D.~M.}\ \bibnamefont {van Egmond}}, \bibinfo {author} {\bibfnamefont {U.}~\bibnamefont {Reinosa}}, \bibinfo {author} {\bibfnamefont {J.}~\bibnamefont {Serreau}},\ and\ \bibinfo {author} {\bibfnamefont {M.}~\bibnamefont {Tissier}},\ }\href {https://doi.org/10.21468/SciPostPhys.12.3.087} {\bibfield  {journal} {\bibinfo  {journal} {SciPost Phys.}\ }\textbf {\bibinfo {volume} {12}},\ \bibinfo {pages} {087} (\bibinfo {year} {2022})}\BibitemShut {NoStop}%
\bibitem [{\citenamefont {van Egmond}\ \emph {et~al.}(2024)\citenamefont {van Egmond}, \citenamefont {Oliveira}, \citenamefont {Reinosa}, \citenamefont {Serreau}, \citenamefont {Silva},\ and\ \citenamefont {Tissier}}]{vanEgmond2024TheLattice}%
  \BibitemOpen
  \bibfield  {author} {\bibinfo {author} {\bibfnamefont {D.~M.}\ \bibnamefont {van Egmond}}, \bibinfo {author} {\bibfnamefont {O.}~\bibnamefont {Oliveira}}, \bibinfo {author} {\bibfnamefont {U.}~\bibnamefont {Reinosa}}, \bibinfo {author} {\bibfnamefont {J.}~\bibnamefont {Serreau}}, \bibinfo {author} {\bibfnamefont {P.~J.}\ \bibnamefont {Silva}},\ and\ \bibinfo {author} {\bibfnamefont {M.}~\bibnamefont {Tissier}},\ }\href {http://arxiv.org/abs/2412.07930} {\bibinfo {title} {{The center-symmetric Landau gauge meets the lattice}}} (\bibinfo {year} {2024}),\ \Eprint {https://arxiv.org/abs/2412.07930} {arXiv:2412.07930 [hep-lat]} \BibitemShut {NoStop}%
\bibitem [{\citenamefont {van Egmond}\ and\ \citenamefont {Reinosa}(2023)}]{vanEgmond2023GaugeSymmetries}%
  \BibitemOpen
  \bibfield  {author} {\bibinfo {author} {\bibfnamefont {D.~M.}\ \bibnamefont {van Egmond}}\ and\ \bibinfo {author} {\bibfnamefont {U.}~\bibnamefont {Reinosa}},\ }\href {https://doi.org/10.1103/PhysRevD.108.054029} {\bibfield  {journal} {\bibinfo  {journal} {Phys. Rev. D}\ }\textbf {\bibinfo {volume} {108}},\ \bibinfo {pages} {054029} (\bibinfo {year} {2023})}\BibitemShut {NoStop}%
\bibitem [{\citenamefont {van Egmond}\ and\ \citenamefont {Reinosa}(2024)}]{vanEgmond2024Center-symmetricConfinement}%
  \BibitemOpen
  \bibfield  {author} {\bibinfo {author} {\bibfnamefont {D.~M.}\ \bibnamefont {van Egmond}}\ and\ \bibinfo {author} {\bibfnamefont {U.}~\bibnamefont {Reinosa}},\ }\href {https://doi.org/10.1103/PhysRevD.109.036002} {\bibfield  {journal} {\bibinfo  {journal} {Phys. Rev. D}\ }\textbf {\bibinfo {volume} {109}},\ \bibinfo {pages} {036002} (\bibinfo {year} {2024})}\BibitemShut {NoStop}%
\bibitem [{\citenamefont {Mari~Surkau}\ and\ \citenamefont {Reinosa}(2024)}]{MariSurkau2024DeconfinementDependences}%
  \BibitemOpen
  \bibfield  {author} {\bibinfo {author} {\bibfnamefont {T.}~\bibnamefont {Mari~Surkau}}\ and\ \bibinfo {author} {\bibfnamefont {U.}~\bibnamefont {Reinosa}},\ }\href {https://doi.org/10.1103/PhysRevD.109.094033} {\bibfield  {journal} {\bibinfo  {journal} {Phys. Rev. D}\ }\textbf {\bibinfo {volume} {109}},\ \bibinfo {pages} {094033} (\bibinfo {year} {2024})}\BibitemShut {NoStop}%
\bibitem [{\citenamefont {Fromm}\ \emph {et~al.}(2012)\citenamefont {Fromm}, \citenamefont {Langelage}, \citenamefont {Lottini},\ and\ \citenamefont {Philipsen}}]{Fromm2012ThePotentials}%
  \BibitemOpen
  \bibfield  {author} {\bibinfo {author} {\bibfnamefont {M.}~\bibnamefont {Fromm}}, \bibinfo {author} {\bibfnamefont {J.}~\bibnamefont {Langelage}}, \bibinfo {author} {\bibfnamefont {S.}~\bibnamefont {Lottini}},\ and\ \bibinfo {author} {\bibfnamefont {O.}~\bibnamefont {Philipsen}},\ }\href {https://doi.org/10.1007/JHEP01(2012)042} {\bibfield  {journal} {\bibinfo  {journal} {J. High Energy Phys.}\ }\textbf {\bibinfo {volume} {2012}}\bibinfo  {number} { (1)},\ \bibinfo {pages} {42}}\BibitemShut {NoStop}%
\bibitem [{\citenamefont {Brown}\ \emph {et~al.}(1990)\citenamefont {Brown}, \citenamefont {Butler}, \citenamefont {Chen}, \citenamefont {Christ}, \citenamefont {Dong}, \citenamefont {Schaffer}, \citenamefont {Unger},\ and\ \citenamefont {Vaccarino}}]{Brown1990OnQuarks}%
  \BibitemOpen
\bibfield  {number} {  }\bibfield  {author} {\bibinfo {author} {\bibfnamefont {F.~R.}\ \bibnamefont {Brown}}, \bibinfo {author} {\bibfnamefont {F.~P.}\ \bibnamefont {Butler}}, \bibinfo {author} {\bibfnamefont {H.}~\bibnamefont {Chen}}, \bibinfo {author} {\bibfnamefont {N.~H.}\ \bibnamefont {Christ}}, \bibinfo {author} {\bibfnamefont {Z.}~\bibnamefont {Dong}}, \bibinfo {author} {\bibfnamefont {W.}~\bibnamefont {Schaffer}}, \bibinfo {author} {\bibfnamefont {L.~I.}\ \bibnamefont {Unger}},\ and\ \bibinfo {author} {\bibfnamefont {A.}~\bibnamefont {Vaccarino}},\ }\href {https://doi.org/10.1103/PhysRevLett.65.2491} {\bibfield  {journal} {\bibinfo  {journal} {Phys. Rev. Lett.}\ }\textbf {\bibinfo {volume} {65}},\ \bibinfo {pages} {2491} (\bibinfo {year} {1990})}\BibitemShut {NoStop}%
\bibitem [{\citenamefont {Zwanziger}(1989)}]{Zwanziger1989LocalHorizon}%
  \BibitemOpen
  \bibfield  {author} {\bibinfo {author} {\bibfnamefont {D.}~\bibnamefont {Zwanziger}},\ }\href {https://doi.org/10.1016/0550-3213(89)90122-3} {\bibfield  {journal} {\bibinfo  {journal} {Nucl. Phys. B}\ }\textbf {\bibinfo {volume} {323}},\ \bibinfo {pages} {513} (\bibinfo {year} {1989})}\BibitemShut {NoStop}%
\bibitem [{\citenamefont {Dudal}\ \emph {et~al.}(2008)\citenamefont {Dudal}, \citenamefont {Gracey}, \citenamefont {Sorella}, \citenamefont {Vandersickel},\ and\ \citenamefont {Verschelde}}]{Dudal2008RefinementResults}%
  \BibitemOpen
  \bibfield  {author} {\bibinfo {author} {\bibfnamefont {D.}~\bibnamefont {Dudal}}, \bibinfo {author} {\bibfnamefont {J.~A.}\ \bibnamefont {Gracey}}, \bibinfo {author} {\bibfnamefont {S.~P.}\ \bibnamefont {Sorella}}, \bibinfo {author} {\bibfnamefont {N.}~\bibnamefont {Vandersickel}},\ and\ \bibinfo {author} {\bibfnamefont {H.}~\bibnamefont {Verschelde}},\ }\href {https://doi.org/10.1103/PhysRevD.78.065047} {\bibfield  {journal} {\bibinfo  {journal} {Phys. Rev. D}\ }\textbf {\bibinfo {volume} {78}},\ \bibinfo {pages} {065047} (\bibinfo {year} {2008})}\BibitemShut {NoStop}%
\bibitem [{\citenamefont {Maas}(2009)}]{Maas2008MoreTheory}%
  \BibitemOpen
  \bibfield  {author} {\bibinfo {author} {\bibfnamefont {A.}~\bibnamefont {Maas}},\ }\href {https://doi.org/10.1103/PhysRevD.79.014505} {\bibfield  {journal} {\bibinfo  {journal} {Phys. Rev. D}\ }\textbf {\bibinfo {volume} {79}},\ \bibinfo {pages} {014505} (\bibinfo {year} {2009})}\BibitemShut {NoStop}%
\bibitem [{\citenamefont {Maas}(2011{\natexlab{a}})}]{Maas2011OnOrbit}%
  \BibitemOpen
  \bibfield  {author} {\bibinfo {author} {\bibfnamefont {A.}~\bibnamefont {Maas}},\ }\href {https://doi.org/10.22323/1.136.0028} {\bibfield  {journal} {\bibinfo  {journal} {PoS}\ }\textbf {\bibinfo {volume} {QCD-TNT-II}},\ \bibinfo {pages} {028} (\bibinfo {year} {2011}{\natexlab{a}})}\BibitemShut {NoStop}%
\bibitem [{\citenamefont {Maas}(2011{\natexlab{b}})}]{Maas2011PropertiesOrbits}%
  \BibitemOpen
  \bibfield  {author} {\bibinfo {author} {\bibfnamefont {A.}~\bibnamefont {Maas}},\ }in\ \href {https://doi.org/10.22323/1.105.0279} {\emph {\bibinfo {booktitle} {Proceedings of The XXVIII International Symposium on Lattice Field Theory {\textemdash} PoS(Lattice 2010)}}},\ Vol.\ \bibinfo {volume} {105}\ (\bibinfo {year} {2011})\ p.\ \bibinfo {pages} {279}\BibitemShut {NoStop}%
\bibitem [{\citenamefont {Maelger}\ \emph {et~al.}(2018{\natexlab{b}})\citenamefont {Maelger}, \citenamefont {Reinosa},\ and\ \citenamefont {Serreau}}]{Maelger2018UniversalQuarks}%
  \BibitemOpen
  \bibfield  {author} {\bibinfo {author} {\bibfnamefont {J.}~\bibnamefont {Maelger}}, \bibinfo {author} {\bibfnamefont {U.}~\bibnamefont {Reinosa}},\ and\ \bibinfo {author} {\bibfnamefont {J.}~\bibnamefont {Serreau}},\ }\href {https://doi.org/10.1103/PhysRevD.98.094020} {\bibfield  {journal} {\bibinfo  {journal} {Phys. Rev. D}\ }\textbf {\bibinfo {volume} {98}},\ \bibinfo {pages} {094020} (\bibinfo {year} {2018}{\natexlab{b}})}\BibitemShut {NoStop}%
\bibitem [{\citenamefont {Reinosa}\ \emph {et~al.}(2015{\natexlab{c}})\citenamefont {Reinosa}, \citenamefont {Serreau},\ and\ \citenamefont {Tissier}}]{Reinosa:2015oua}%
  \BibitemOpen
  \bibfield  {author} {\bibinfo {author} {\bibfnamefont {U.}~\bibnamefont {Reinosa}}, \bibinfo {author} {\bibfnamefont {J.}~\bibnamefont {Serreau}},\ and\ \bibinfo {author} {\bibfnamefont {M.}~\bibnamefont {Tissier}},\ }\href {https://doi.org/10.1103/PhysRevD.92.025021} {\bibfield  {journal} {\bibinfo  {journal} {Phys. Rev. D}\ }\textbf {\bibinfo {volume} {92}},\ \bibinfo {pages} {025021} (\bibinfo {year} {2015}{\natexlab{c}})}\BibitemShut {NoStop}%
\end{thebibliography}%
\end{document}